\documentclass[10pt,twoside]{article}
\usepackage{amssymb}
\usepackage{amsmath}
\usepackage{epsfig}
\usepackage{color}

\font\fourbf=cmbx14

\font\tenrm=cmr9

\font\tenit=cmti9

\newcommand{\lle}{\mbox{$\langle$}}
\newcommand{\rle}{\mbox{$\rangle$}}

\newcommand{\bfu}{\mbox{\boldmath$\bf u$}}
\newcommand{\bfv}{\mbox{\boldmath$\bf v$}}
\newcommand{\bfx}{\mbox{\boldmath$\bf x$}}
\newcommand{\bfy}{\mbox{\boldmath$\bf y$}}

\newcommand{\bfal}{\mbox{\boldmath$\alpha$}}

\newcommand{\BB}{\begin{equation}}
\newcommand{\EE}{\end{equation}}

\newcommand{\BBEQ}{\begin{eqnarray}}
\newcommand{\EEEQ}{\end{eqnarray}}
\begin{document}

\vspace{18pt}
 \centerline{\fourbf
Modeling of random bimodal structures of }


\vspace{2pt}

 \centerline{\fourbf composites (application to solid propellants):}
 \vspace{2pt}

 \centerline{\fourbf  I. Simulation of random packs}

\vspace{6pt}

\vspace{10pt}  \centerline{\bf V.A. Buryachenko$^{1,2)}$\footnote{\noindent \tenrm
$^{1)}$Current address: Micromechanics \& Composites LLC, 2520 Hingham Lane, Dayton, OH 45459, USA\\
$^{2)}$IllinoisRocstar LLC, 60 Hazelwood Drive, Champaign, IL 61820, USA\\
$^{3)}$ Department of Aerospace Engineering, University of Illinois, Urbana-Champaign, IL 61820, USA
}, T.L. Jackson$^{2,3)}$, G. Amadio$^{3)}$}

\vspace{2pt}


\vspace{4pt}


\noindent {\baselineskip=9pt
{\bf Abstract:}
We consider a composite medium, which consists of a homogeneous matrix
containing a statistically homogeneous set of multimodal spherical inclusions.
This model is used to represent the morphology of heterogeneous solid propellants (HSP) that are
widely used in the rocket industry.
The Lubachevsky-Stillinger algorithm is used to generate morphological models of HSP
with large polydisperse packs of spherical inclusions.
We modify the algorithm by proposing a random shaking procedure that
leads to the stabilization of a statistical distribution
of the simulated structure that is homogeneous, highly mixed, and protocol
independent (in sense that the statistical parameters estimated do not depend on the
basic {\color{black}simulation} algorithm).
Increasing the number of shaking has a twofold effect.
First, the system becomes more homogeneous and well-mixed. Second,
the stochastic fluctuations of statistical parameters (such as e.g.
radial distribution function, RDF), estimated by averaging of these structures,
tend to diminish.}
\vspace{-3pt}

\noindent
{\bf Keywords:}  Microstructures,  random packing, spherical particles, inhomogeneous material.

\smallskip

\medskip
\noindent{\bf 1. Introduction}
\medskip

There is a growing recognition within the nano- and micromechanics community
that the prediction of mechanical properties of heterogeneous materials, such as solid
propellants,
must take into account the various scales (such as the particle sizes) and cannot be treated
as a homogeneous material.
This is due in large part by the use
of image {\color{black}analysis} and computer-simulation methods on one hand and advanced
experimental techniques (such as X-ray tomography and electron microscopy)
and improved materials processing
(prescribed structure controlled by processing) on the other.
The prediction of the behavior of composite materials in terms of the mechanical
properties of constituents and their microstructure is a central problem of micromechanics, while
the quantitative description of the microtopology of heterogeneous media
is crucial in the prediction of overall mechanical and physical properties of these materials.
After many years of comprehensive study by direct measurements and empirical relations
that is extremely laborious, the structure of microinhomogeneous
materials is not completely understood.
Influence of microstructure on mechanical properties of
microinhomogeneous media drastically increases for
highly packed particulate media such as, e.g., concentrated
suspensions (see, e.g., Chong, Christiansen and Baer, 1971), concrete (see for references
Garboczi and Bentz, 1997; Amparano, Xib and Roh, 2000), dental restorative
composite materials (see Lingois and Berglund, 2002; Wissler {\it et al.}, 2003;
Tanimoto {\it et al.}, 2004), and solid propellants.

Heterogeneous solid propellants (HSPs) commonly used in aerospace propulsion are likely to play an
important role for as long as rockets are {\color{black}built}.
Fundamentally, they consist of multimodal spherical particles of solid oxidizer
with extremely high volume fractions ($>$0.9), typically ammonium perchlorate,
dispersed in a rubbery fuel binder fuel such as hydroxi--terminated--polybutadiene (HTPB) or
polybutadiene acrylonitrile (PBAN). Designers of rockets are concerned with a
number of propellant--related issues, including the burning rate, the thermal and
mechanical properties of the propellants, and, for metallized propellants, the
behavior of the metal particles at the surface, including agglomeration.
(see Massa, Jackson and Short, 2003; Webb and Davis, 2006; Matou\v{s} and Geubelle, 2006; Matou\v{s} {\color{black}{\it et al.}}, 2007;
Maggi {\it et al.}, 2008;  {\color{black}Xu, Aravas and Sofronis,} 2008).

The prediction of effective mechanical properties of polymer-bonded explosives
provides a new challenge. These particulate composites contain extremely high volume
fractions of explosive particles ($>$0.9) in a polymer binder typically exhibiting
viscoelastic behavior (see, e.g., Banerjee and Adams, 2004). In explosives, particle
sizes range from a few $\mu$m for oxidizer particles in the ``dirty binder"
to 10s of $\mu$m for the metal flakes and 100s of $\mu$m for the largest oxidizer crystals.
The oxidizer (HMX) particle shapes are more polyhedral than spherical
(see Stafford and Jackson, 2010). Large dynamic deformation during processing and
shock loading can fracture the microstructure and microcracks can
create additional internal surfaces or porosity that
greatly influences explosive initiation sensitivity (see e.g. Baer {\it et al.}, 2007).

In many important cases of practical interest (e.g. solid propellants),
the shape of the heterogeneities can be assumed as a spherical.
Computer simulation of topologically disordered structures by the random packing of hard
spherical particles in 2-$D$ and 3-$D$ cases has a long history originated in the theory of liquids.
This problem is closely connected with the known fundamental problem of statistical physics
- description of the behavior of the particle system with
interaction potential of hard spheres (see e.g Binder and Heermann, 1997). Random packing of spheres
has been studied very extensively due to their technological importance, and their
{\color{black}capability of modeling} the simple liquid, concentrated suspensions, amorphous
and {\color{black}powder} materials.

Computer simulation of packing problems can be classified into three
groups of methods: molecular dynamic, Monte Carlo, and dense random packing.
Much progress in the theory of the dense random packing was {\color{black}achieved} by
the use of two kinds of methods: the sequential generation models and the
collective rearrangement models. In the sequential model with the fixed radii of particles proposed
by Bennet (1972) [see also the modified algorithms Boudreaux and Gregor, 1977; Lu, Ti, and Ishizaki, 1994; Kansal, Truskett and Torquato (2000)], so called cluster growth model, a particle being added
to the surface of particle cluster which grows outwards is placed sequentially to the
point closest to the original such that the new particle established contact with
three existing spheres in the cluster.
The key geometric procedure by Jerier {\it et al.} (2010) consists in placing
one sphere in contact with four non-coplanar spheres.
In the second type of sequential generation model,
called the model of ``rigid sphere free fall into a virtual box",
one particle is dropped vertically each time from the random point
onto the surface of an existing particle cluster growing upwards
(see e.g. Nolan and Kavanagh, 1992; Cesarano, McEuen and Swiler 1995; Kondrachuk, Shapovalov and Kartuzov, 1997, Furukawa, Imai and Kurashige, 2000; Kanuparthi, {\it et al.}, 2009).
For contrast, dynamic methods assume the reorganization
of whole packing due to either short or long range interactions between particles.
In the collective rearrangement model (CRM, or concurrent), $N$ points randomly distributed in a virtual box
are assigned both radii and random motion vectors. Each sphere is moved
until there are no overlaps. Then the radii are increased and the process is repeated until any
further increase in radii or any displacement of the spheres create overlaps that can not be eliminated
[the different versions of this method can be found e.g. in
Clarke and Willey, 1987; Lubachevsky and Stillinger, 1990; Lubachevsky, Stillinger and Pinson, 1991;
Zinchenko, 1994; Knott, Jackson and Buckmaster, 2001; Kochevets {\it et al.}, 2001; He and Ekere, 2001; see also {\color{black}Oger} {\it et al.}, 1998
where a detailed description of the advantages and drawbacks of the different algorithms were presented].
 Classical molecular dynamics can also be
considered as a CRM, where particle size is fixed. The CRMs satisfactorily reproduce
the real packing properties, but require high time-consuming
computation. More recently, numerical simulations were performed to realize
homogeneous and isotropic packing of spheres by various methods, for instance,
by assuming hypothetical spheres having dual structure whose
inner diameter defines the true density and the outer one a nominal
density (see Jodrey and Tory, 1985). We only consider the spherical particles because only a
single parameter, the radius, defines a single possible type of contact among
particles, which can be estimated analytically and easily.

The a priori prediction of the maximum packing fraction, $c_m$, for a system of
particles is still an open question, despite much study.
Packing densities for close lattice packing of monomodal particles are
$\pi/\sqrt{12}\approx 0.9069$
(triangular) in the case of disks packing into the plane, and $\pi/\sqrt{18}\approx 0.7405$
(fcc or hcp) in the case of spheres packed into $R^3$.
 However, a well-mixed random packing does not
self-assemble into one of the periodic arrangements but instead forms a so-called
random close-packed (RCP) arrangement which is not well defined, protocol dependent (in both the numerical simulation and experimental sense, see
McGeary, 1961; Berryman 1983; Cheng, Guo and Lay, 2000; Torquato, Truskett and Debenetti, 2000; Aste,
 Saadatfar and Senden, 2005), and can vary in the range $0.59\doteqdot 0.64$ for 3D unimodal spheres. Here the low value corresponds to the loose packs (LRP), while the upper value is a characteristic of so-called dense packs (DRP). A shaking process makes it possible to unlock the spheres from
the jamming configuration and allows them to find the most homogeneous and mixed arrangement.
So, Nowak {\it et al.} (1997) demonstrated by an electromagnetic vibration exciter that under vibrations the bead packing evolves from an initial, low-density configuration towards higher density.
The measured densities are extrapolated to eliminate finite-size effect that enables one to estimate the DRP's density $c_m=0.6366\pm 0.004$ [Stoyan (1998) proposed elegant hypothesis for $c_m=2/\pi\approx 0.63662$].
Various algorithms have been devised to simulate reordering due to shaking or vibration of dense packing (see, e.g., Barker, 1993; Lee {\it et al.}, 2009) which reduces the volume concentration of the high density jam configuration.

For multimodal systems, small spheres may be placed into the spaces between packed large spheres.
For the infinitely different sizes of packed spheres, the maximum $c_m=0.637\cdot{\color{black}(2-0.637)}=0.868$.
The volume fraction ratio $\xi$ between
large and small spheres is also important, with maximum packing obtained at about
$\xi=0.60\doteqdot 0.75$ large particles (Shapiro and Probstein 1992; Maggi, {\it et al.}, 2008, 2010).
So, for the packs with size ratio $\lambda=6.5:1$ Maggi {\it et al.} (2008) simulated the packing fractions of
$0.687$ and $0.731$ for the {\color{black}coarse}--to--fine volume ratio $\xi=0.90$ and $\xi=0.75$, respectively.
Trimodal, multimodal, and polydisperse systems can reach even higher packing fractions
(Zou {\it et al.}, 2003).
However, packing configurations containing a wide range of inclusion concentrations have been investigated less
and invite further investigation.

The paper is organized as follow.
In Section 2, the quantitative descriptors of the bimodal spherical dispersion are discussed
in order to describe the pattern of particle location as it really exists rather than as
described by some assumed model.
Since random packing structures are strongly dependent on the procedure
of their generation, in Section 3 we consider a popular algorithm of the collective rearrangement model (CRM) and its combination with the shaking procedure (shaking collective rearrangement model, SCRM) adapted for obtaining the most homogeneous configurations which are protocol independent.
In Section 4 we recall the basic concepts defining some classical method of micromechanics and present explicit formula for both effective elastic moduli and stress concentrator factor for composites with bimodal distribution of spherical particles.
In Section 5 one estimates the RDFs of configurations generated by the CRM and SCRM for unimodal and bimodal distributions of spherical particles. Diminishing of stochastic fluctuations of the RDFs are reached by using of parallel computing. Comparison of these simulations for unimodal packs with the real scanned data obtained by X-ray tomography demonstrates an advantage of the SCRM with respect to CRM. One performs a detailed analysis of local maxima and minima of RDFs for bimodal simulated packs.

Estimated RDFs are used in an accompanying paper by Buryachenko (2012) for evaluation of both the effective elastic moduli and stress concentrator factors of local stresses in linear elastic composites, which  consist of a homogeneous matrix containing a statistically homogeneous set of bimodal spherical inclusions.

\medskip
\noindent{\bf 2 Statistical description of random structure particulate composites}
\medskip


Let a linear elastic infinite body $R^d$
{contains} an open bounded domain $w\subset R^d$ (window of observation)
with a boundary $\Gamma$ and with an indicator
function $W$ and space dimensionality $d$
($d=2$ and $d=3$ for 2-$D$ and 3-$D$ problems, respectively).
The domain $w$ contains a homogeneous matrix $v^{(0)}$ and
a random finite set $X=(v_i)$ $(i=1,\ldots, N(w))$ of inclusions
$v_i$ with centers $\bfx_i$ and with characteristic
functions $V_i(\bfy)$ equals one {\color{black}if} $\bfy\in v_i$ and zero otherwise
and bounded by the spherical surfaces
$\Gamma_i^{(1)}=\{\bfy:\ |\bfy-\bfx_i|=a^{(1)}\}$ and $\Gamma_i^{(2)}=\{\bfy:\ |\bfy-\bfx_i|=a^{(2)}\}$ of two radii
$a^{(1)}$ and $a^{(2)}$, respectively; $v^{(k)}=\cup v_i^{(k)}$ $(i=1,2,\ldots,N^{(k)}, \ N^{(1)}+N^{(2)}=N, \ k=1,2)$.
$N=N(w)$ denotes the random number of points $\bfx_i$ falling in $w$;
$\hat N(w)$ is the observed number of points in $w$.

At first we summarized some basic ideas, notations, and quantities for random point processes
such as the centers of particles (for more details see Ripley, 1977, 1981: Diggle, 2003;
K\"onig, 1991; Stoyan and Stoyan, 1994; Stoyan, 2000, Torquato, 2002).
No confusion will arise below in using of terminologies from the different scientific subfields.
So, in micromechanics one uses the word ``field" where statisticians would prefer to speak of �process�
although typically there is no time-dependence considered (see, e.g., Stoyan, 2000).
We will consider statistically homogeneous
(or stationary) and isotropic ergodic
random process $X$, keeping in mind that the stationary, isotropy, and ergodicity can never be tested statistically
in their full generality. Stationarity means invariance under arbitrary translation, and isotropy
means invariant under arbitrary rotation. The ergodicity ensured that one sample (one point
pattern) is sufficient for obtaining statistically secure results, assuming the convergence of results obtained
for infinitely expanding observation window $w$.
The intensity, $n=EX([0,1]^d)$, of a stationary point process is
the mean number of points in the unit cube. A standard estimator of the intensity is
$\hat n=N(w)/ \bar w$, $\bar w\equiv {\rm mes}\ w$.

The packed random structure can be characterized by several parameters,
such as packing density, coordination number, radial distribution function,
particle cage, inter-particle spacing, and others.
The various methods of estimating the effective properties of a composite material
use a knowledge of
statistical geometrical information about the microstructure.
In order to incorporate the spatial arrangement of components in a micromechanical
simulation, it is essential to quantitatively characterize the
random structure of the composite. The most important factors
characterizing the microstructure of composite material are the shape,
volume fraction, and arrangement (random or regular)
of the components that permit the calculation of bounds of effective moduli.

The popular statistical description of microinhomogeneous media
is based on expectations of products of the characteristic function
$V^{(i)}(\bfy)$, assuming that the role of the matrix is assigned to phase `0'.
So, improved bounds on a variety of different effective
properties have been derived in terms of $n$-point probability density
\BB
S_n^{(i)}(\bfy^n)=\langle V_i(\bfy_1),\ldots V_i(\bfy_n)\rangle,
\EE
 where the angle brackets $\langle(\cdot)\rangle$ denote an ensemble average; i.e., the probability of simultaneously finding of
$n$ points in a specified geometrical arrangement $\bfy^n
\equiv \bfy_1,\ldots,\bfy_n$ in one of the phases. In particular, two-point probability density is widely used for statistical description of the HSPs (see, e.g.,  Matou\v{s} and Geubelle, 2006; Maggi {\it et al.}, 2008;  Lee, Gillman and Matou\v{s}, 2011)

An alternative related approach of quantitative description of
the composite microstructure is based on the consideration of the inclusion centers
statistically described by the multiparticle probability densities
$f_m(\bfx_1,\ldots,\bfx_m)$ that give the probability $f_m(\bfx_1,\ldots,\bfx_m)
d\bfx_1\ldots d\bfx_m$ to find a sphere center in the vicinities $d\bfx_1\ldots d\bfx_m$
of the point $\bfx^m=(\bfx_1,\ldots,\bfx_m)$. {The $f_m$ are the most basic descriptors
that characterize the structure of many-particle system and have been well-studied in the statistical mechanics of liquids
(Hansen and McDonald, 1986).}
In particular, $f_1=n$, where $n$ is the number density of inclusions connected
with the volume fraction $c=n\bar v_i$, $\bar v_i\equiv{\rm mes} v_i$.
For the statistically homogeneous isotropic media in the framework of the simplest ``two-point" level,
Markov and Willis (1998) demonstrated a simple interconnection expressing
$S_2^{(1)}(\bfy^2)$ as a simple one-tuple integral containing the two-point probability density $f_2(r)$ ($r=|\bfx_1-\bfx_2|$).
 Torquato and Stell (1985) have related $S^{(1)}_n$ to multidimensional
integrals over the infinite set of $m$-particle densities $f_1,\ldots f_m$
($m\to \infty$).

The widely used informative function that describes the point distribution is the
{\it second-order intensity function} $K(r)$ (called also Ripley's (1977) $K$ function) defined as the number of further points
expected to be located within a distance $r$ of an arbitrary point
divided by the number of points per unit area $n$ (Ripley, 1977; Diggle, 1983; see also Pyrz, 2004; Silberschmidt, 2008). Since points lying outside the observation window $w$ are
nonobserved, the latter depends strongly on the shape and the size of $w$.
It is our aim to simulate a typical realization that also includes interaction to the structure element
outside of the window while avoiding systematic errors or biases in the estimation procedure.
The effect of the edge of the domain $w$ becomes increasingly dominant at the dimensional
increases. A number of special edge corrections are known. A naive way proposed in
{\it the minus-sampling} method is to consider $w^*$ within domain $w$ and allow measurements
from an object in $w^*$ to an object in $w$. Although the effective sample size is then the number of points
in $w$, the method, of course, leads to a big loss of information.
A much better idea of edge-correction of the estimator for $K(r)$
was suggested by {\color{black}Ripley} (1977)
\BB
\hat K(r)=\frac{\bar w}{\hat N^2}\sum_i \sum_{j\not =i}w_{ij}^{-1}I_{ij}(r_{ij}), 
\EE
where $\hat N$ is the number of points $\bfx_k$ ($k=1,\ldots \hat N$) in the field of
observation $w$ with the area $\bar w$,
$I_{ij}(r_{ij})$ is the indicator function equals 1 if $r_{ij}\le r$ and zero otherwise
where $r_{ij}$ is a distance between the typical fixed points $\bfx_i$ and $\bfx_j$.
The term ``typical
point" was used by Stoyan and Stoyan (1994) as a synonym of ``arbitrarily chosen point", where
``arbitrarily" means that a selection scheme is used in which every point has the same chance to be
selected. The typical point concept is a simplified notion of the Palm distribution referring to individual points in a point process (see for details Illian {\it et al.}, 2008).
$w_{ij}$ is the ratio of the circumference contained within $w$ to the whole circumference with radius $r_{ij}$.
If the
entire circle with radius $r$ is placed within the window, $w_{ij}=1$ and it is
smaller than unity otherwise.
For circles intersecting the boundary $\partial w$, the function $w_{ij}$
compensating for the boundedness of $w$ is less than one and has an explicit formula when the field of observation $w$
is rectangular (see Diggle, 1983; the formula and algorithm for 3-$D$ case were proposed by
K\"onig {\it et al.}, 1991). The function $K(r)$ (2) is obtained by averaging over all inclusions at each value of $r$.
Equation (2) is an approximately unbiased estimator,
which is free of systematic errors, for sufficiently small $r$ because $N/\bar w$ is a slightly biased estimator for $n$. Illian {\it et al.} (2008) considered different improvements for the quality of estimations of $K(r)$ and $n$.
In 2-$D$, Diggle (1983) recommended an upper limit of $r$ equal to half the length of the diagonal of a square sampling region.

In a packing simulation only a relatively small number of particles can be considered. To avoid the (usually)
undesired artificial effects of particles which are not surrounded by neighboring particles (near the window of observation boundary) in
all directions, one introduces periodic boundary conditions (see for details Gajdo\v{s}\'ik, and Zeman,  Sejnoha, 2006; Steinhauser and Hiermaier, 2009).
In such a case, an alternative {\it toroidal edge correction} (see Ripley, 1979) is often used
in which each 2D rectangular box $w$ can be regarded as a torus,
so that points on opposite edges are considered to be closed (a 3D case is analyzed in a similar manner). Then $w$ can be considered to be part of a grid of
identical boxes, forming a border around the pattern inside $w$. If a moving particle leaves the
central simulation box, then one of its image particles enters the central box from the opposite direction. Each of the image particles in the neighboring boxes
(the numbers of these neighboring boxes equal 8 and 26 in 2D and 3D case, respectively) moves in exactly the same way.
 Precisely this method
will be explored hereafter in this paper for the elimination of the boundary effect.
So, we start with a cube $w=\{[-1,1]^d\}$ with the center ${\bf 0}\in R^d$ and its periodically located neighbors
labeled by the multiplet of integer numbers $\bfal=(\alpha_1,\ldots,\alpha_d)
\in Z^+$ where $\alpha_i=0,\pm 1$ ($i=1,\ldots,d)$. $N$ random points $\bfx_i$
($i=1,\ldots,N)$ in the central cube $w$ are periodically reflected into
the neighboring cubes $w^{\small \bfal}$.
Distances are then measured taking into account the minimum image convention according to which the real distance between any
two particles is given by the shortest distance of any of their images
 in the surrounding periodic cubes $w^{\small \bfal}$ (see
{\color{black}Ripley}, 1981): $|\bfx_i-\bfx_j|=\min_{\small \bfal} |\bfx_i-\bfx_j^{\small \bfal}|=\min_{\small \bfal}|\bfx_i-\bfx_j-2(\alpha_1+\ldots+\alpha_d)|$ where $\bfx_j^{\small \bfal}\in w^{\small\bfal}$ is a periodic mapping of $\bfx_j\in w$ into $w^{\small\bfal}$.
Toroidal edge correction allows us to analyze all $N$ particles in the central box $w$ without using the minus-sampling method because even for the points
$\bfx_i$ placed near the boundary of $w$ all surrounding points are visible.
An alternative approach (eliminating the boundary effect of the virtual box $w$) is based on the use of spherical
boundary conditions instead of periodic ones (see e.g. Hall, 1988; Tobochnik and Chapin, 1988). There one
simulates hard disks (more exactly a circular cap which can be visualized as a contact lens on the surface of
an eyeball) on the surface of the ordinary three-dimensional sphere and hard spheres
on the ``surface" of a
four-dimensional {\color{black}hypersphere}. The advantage of this procedure is that there is no preferential direction, and
it is impossible to pack particles into perfect regular periodic configurations.
The considered both toroidal and spherical edge corrections hold for simulated structures rather than for the real observation window $w$ obtained by visual cutting out from a large sample. In the last case, either the edge correction (2) or minus-sampling method should be used.

For unimodal distributions, the widely used two-point density $f_2(r)=f_2(\bfx_1,\bfx_2)$ ($r=|\bfx_1-\bfx_2|$)
is expressed in terms of {\it radial distribution function} $g(r)$ (RDF) (called also the pair correlation function, see Stoyan and Stoyan, 1994) as
\BB
f_2(r)=n^2g(r),
\EE
which is a function of the inter-point distance $r$.
{\color{black}Hence}, $ng(r)\bar{s}_d dr$ is a probability to find a point whose center lies in an infinitesimal
 spherical shell $s_d\subset R^d$ of inner radius $r$ with a surface $\bar {s}_d$ ($\bar {s}_d=2\pi r, 4\pi r^2$ for $d=2,3$, respectively)
thickness $dr$ and centered in a specific particle.
RDF $g(r)$ (3) which plays a role similar to the
variance in a classical analysis of random variables is defined as the radial
distribution of the average number
of sphere centers per unit area in a spherical shell. The RDF can be estimated
from second-order intensity function $K(r)$ as (see e.g., Stoyan and Stoyan, 1994; Pyrz, 2004)
\BB
g(r)={1\over d\omega_d r^{d-1}}{d K(r)\over dr}, 
\EE
where $\omega_d$ is the volume of the unit sphere in $R^d$ ($\omega_d=\pi,\ 4/3\pi$ for $d=2,\ 3$, respectively).
The $K$-function is
seldom used in the study of packing, and the RDF $g(r)$ is used instead despite the fact that the estimation of $g(r)$ is more delicate due to numerical evaluation of the derivative (4).
The RDF is related to the derivative of $K(r)$ (4), and is {\color{black}therefore more} sensitive
to changes in the spatial order than is the function $K(r)$. For a completely random point process (i.e. a homogeneous Poisson process) $K(r)=\omega_d r^d$, $g(r)\equiv 1$. Values of
the RDF $g(r)$ larger than 1 indicate that the interposing distances around $r$ appear
more frequently compared to those in a completely random point process (typically there is
clustering), whereas values of g(r) smaller than 1 indicate that the corresponding distances
are rare, that there is mutual inhibition. RDF takes all values from
zero to large values at $g(r\to a^+)=g(a^+)$ {\color{black}($a^+=\lim_{\epsilon\to 0}a+\epsilon,\ \epsilon>0$)}, and the second neighbor shell at the large $c$ becomes split exhibiting a shoulder. For $r\to \infty$ it tends to 1, and the rate of convergence reflects a degree of randomness of the structure considered in a comparison with a completely random Poisson field. The radius $r_0$ such that $g(r)\approx 1$ at $r>r_0$ determines the range of geometrical disorder where a reference heterogeneity does not distinguish individual neighbors and ``fills" the surrounding as a continuum.

Up to now we only considered unimodal distribution of identical heterogeneities. The introduction of mark correlation function (see Stoyan and Stoyan, 1994;
Gavrikov and Stoyan, 1995; Pyrz, 2004; Rudge, Holness and Smith, 2008) may facilitate the characterization of more complex aspect of random structure. One attaches to each point with a position $\bfx_i$ of the center set a typical feature which may correspond to the size, shape and type of inclusions
served as ``point identifiers". In our case, the ``points" are the particle locations, while
``marks" describe the ``particle type" with reference to the particle radii $a^{(1)}$ and $a^{(2)}$. Then Eq. (2) for the toroidal edge correction
(when $w_{ij}\equiv 1$) can be generalized to the partial $K_{km}$ functions
\BB
 \hat K_{km}(r)={\bar w\over \hat N^{(k)} \hat N^{(m)}}\sum_{i=1}^{N^{(k)}}\sum_{j, j\not =i}^{N^{(m)}}I_{ij}(r_{ij}), 
\EE
where the indexes $i$ and $j$ correspondent to the particles $v^{(k)}$ and $v^{(m)}$ ($k,m=1,2$), respectively.
The interpretation of the partial RDF $g_{km}$ is similar to that of $g(r)$.
It starts with the probability
$f_{2|km}(\bfx_1,\bfx_2)d\bfx_1d\bfx_2$ of
having one of particle type $k$ and the other of type $m$.
Then ($k,m=1,2$)
 \BB
 f_{2|km}(\bfx_1,\bfx_2)=n^{(k)}n^{(m)}g_{km}(r),\ \ \ g_{km}(r)={1\over d\omega_d r^{d-1}}{d \hat K_{km}(r)\over dr}, 
 \EE
 where $n^{(k)}$
 is the intensity of particles $v^{(k)}$, the mean number $k-$points per unit volume, $n=n^{(1)}+n^{(2)}$.
 In accordance with the definition, $f_{2|km}$ is symmetric with respect to $k\leftrightarrow m$ that implies the symmetry
 $g_{km}(r)=g_{mk}(r)$ for $k, m=1,2$ and $\forall r\geq a^{(k)}+a^{(m)}$ (see e.g., Illian {\it et al.}, 2008).
 Introduction of partial descriptors (5) and (6) enables one to describe more accurately the random structures than
 their mass-weighted {\color{black}version} of the spatial repartitions proposed by Gallier (2009) for analysis of solid propellants.

It is also known other types of statistical descriptors (used, e.g., for the bound estimations of affective properties)
 such as point/$q$-particle functions, surface-surface correlation functions,
nearest-neighbor distribution function,
linear-path function, two-point cluster function,
chord-length distribution function
as well as the generalized $n$-point distribution function
for the system of identical spheres
$H_n(\bfy^m;\bfy^{p-m};{\bf r}^q)$, which is defined to be the correlation associated with finding
$m$ points with positions $\bfy^m$ on certain surface within the medium, $p-m$ with positions $\bfy^{p-m}$
in certain space exterior to the spheres, and $q$ sphere centers with positions ${\bf r}^q$, $n=p+q$
(see for details Pyrz, 1994; Torquato, 2002 and references therein).
Although higher-order correlation functions ($N>3$)
are obtained on theoretical grounds, this is not a very practicable approach. However, we will estimate the effective elastic properties
by the MEFM explicitly depending only on the RDFs. Because of this one will consider in details only the RDFs (6).

\bigskip
\noindent{\bf 3 Packing algorithm}
\medskip

\noindent{\textit{\textbf{3.1 Collective rearrangement model}}
\medskip

Collective rearrangement model (CRM) was previously proposed by Lubachevsky, Stillinger, and Pinson
(1991) for the packs of unimodal spheres and generalized by
Knott,
{\color{black} Jackson and Buckmaster}
(2001) and Maggi {\it et al.} (2008) to polydisperse packs (application to heterogeneous solid propellants).
Just for completeness we will briefly reproduce the last version of the CRM accompanied by the shaking procedure
which is {\color{black}adapted} to our goal to generate the most
 homogeneous and highly mixed configurations.

Let $N$ random points $\bfx_i$
($i=1,\ldots,N)$ be located in a central box, periodically reflected into
neighboring boxes. Assign
initial velocities ${\bf u}_i=(u_{i1},\ldots,u_{id})$ to each random point, with
velocity components that are
independently distributed at random between $-1$ and $+1$ and to the
uniformly growing inclusions $v_i\subset v^{(k)}$ ($k=1,2$) with the radii
$a^{(k)}(t)=a^{(k)}_0t$. The centers of the inclusions move according to the equations
\BB
{d\bfx\over dt}={\color{black}{\bf u}_i} 
\EE
with a discontinuous change of the vectors ${\bf u}_i$ at the moment
the particle exits through a face of a central box as well as during collisions
with other inclusions.

The collision time is obtained from the condition that the separation distance is
the current diameter, that for the inclusions $v_i\subset v^{(k)}$ and $v_j\subset v^{(m)}$ ($k,m=1,2$) is
\BB|\bfx_i+{\bf u}_i\Delta t-\bfx_j-{\bf u}_j\Delta t|=(a^{(k)}_0t+a^{(m)}_0t+a^{(k)}_0\Delta t+a^{(m)}_0\Delta t).
\EE
In the case of the collision, the smaller positive root $\Delta t$ of Eq. (8)
defines the collision time $\tau_{ij}$. If the collision takes
place between the inclusions $v_i$ and $v_j$ from central
and neighboring squares (e.g. identified by (-1,1,0)), respectively,
then $\bfx_j$ in Eq. (8) should be replaced by $\bfx_j+(-1,1,0)$ with
a subsequent estimation of
a collision time $\tau_{ij}$. The exit time $\tau_i^{\Gamma}$ is estimated
as the smallest positive time for exiting of the inclusion $v_i$ being considered
through one of {\color{black}the} sides of the central box.
The determination of $\Delta t_*=\min (\tau_{ij},\tau_i^{\Gamma})$ allows {\color{black} us to  estimate} the new inclusion radii $a_0(t+\Delta t_*)$
and identify
either the colliding ($v_{i^*}$ and $v_{j^*}$) or exiting ($v_{i^*}$) inclusions
for which the new velocities and locations are re-estimated. For all other inclusions
the position vectors are updated according to Eq. (7):
$\bfx_i\to \bfx_i+{\bf v}_i\Delta t$.
For these inclusions $\Delta t=\Delta t_*$ and velocities
${\bf v}_i$ remain unchanged. In addition, the collision and exit times
are corrected $\tau_{ij}\to\tau_{ij}-\Delta t_*$ and
$\tau _i^{\Gamma}\to\tau_i^{\Gamma}-\Delta t_*$, respectively.
An optimized cell method with analysis of only collisions between particles in
neighboring cells is used for reduction of the computation time.
The details of the re-estimation of velocities for the inclusions $v_{i^*}$
and $v_{j^*}$ based on the conservation laws of the momentum and energy can be found in
 Knott {\color{black} Jackson and Buckmaster}
 (2000).

\medskip
\noindent{\textit{\textbf{ 3.2 Shaking procedure}}
\medskip

A shaking process (called also Monte-Carlo simulation) gives each particle a small random displacement independent of its neighbors'
positions. This makes it possible to
unlock the particles from the local
jamming configuration
and allows them to find the most homogeneous and mixed arrangement.
Various algorithms have been devised to simulate
reordering due to shaking or vibration (also called Monte Carlo simulation and local-update algorithms, Luijten, 2006)
of dense packing (see Barker, 1993) which reduces the volume concentration of
the high density jam configuration. Packing configurations containing a wide
range of inclusion concentrations have been investigated less and {\color{black} required}
 some additional consideration.
In order to describe the algorithm used in this paper, at first one introduces the following definitions.

{\color{black} For speeding up the calculations, it
is necessary } to carry out the shaking process only within a local shake up window.
The random local shaking of a particle $v_i\subset v^{(k)}$ ($k=1,2$) is established in a shake up window
$w_{\rm sh}^{(k)}=\{\bfx:\ |\bfx-\bfx_i|<R_{\rm sh}^{(k)}a^{(k)}\}$
whereby the selected particle center $\bfx_i$ is randomly moved
to a new position $\bfx'_i$ uniformly distributed in the window $w^{(k)}_{\rm sh}$ with the normalized radius $R_{\rm sh}^{(k)}=$const. ($k=1,2$)
depending on the size of spherical particle in the
multimodal system.
The displacement $\bfx'_i-\bfx_i$ can
be chosen as a randomly oriented vector within a sphere with radius $R^{(k)}_{\rm sh}a^{(k)}$
which permits control over the efficiency of the simulation.
If the particle does not overlap with any other inclusions
the shaking is accepted, otherwise the trial shaking
is repeated until the number of attempted trial shakings exceed some
limit $N_{\rm sh}^{(k)}$. A valid shaking scheme must be ergodic. This means that there is a path in phase space from every state to every other
state, via a succession of trial moves.

Another way to speed up the calculations is to check the
collision partners $v_j$ ($j=j^i_1,\ldots,j^i_{n_t}$) of $v_i$ only in some restricted
neighborhood of $\bfx_i$ called testing window $w_{\rm t}^{(km)}=\{\bfx:\ |\bfx-\bfx_i|<r_{\rm t}^{(km)}\}$
($r_{\rm t}^{(km)}=(1+R_{\rm sh}^{(k)})a^{(k)}+a^{(m)}={\rm const.},\ v_i\subset v^{(k)},\ v_j\subset v^{(m)}$) which was introduced for unimodal particles in 2D case by Buryachenko, {\it et al.} (2003). Thus for bimodal packs of particles $v^{(1)}$ and $v^{(2)}$ we have four different sorts of testing windows $w_{\rm t}^{(km)}$ ($k,m=1,2$).
A similar concept of a near-neighbor list by Donev, Torquato, and Stillinger
(2005) can be considered as a generalization of testing window to the case of nonspherical particles.
In this model, the neighbors of a particle (near-neighbor list, ${\cal N}(i)$) are considered to be the
set of particles whose centers lie
 within some maximum distance called the testing window $w_{\rm t}^i$ of that particle
 rather than the ``geometric neighbors" which can be
determined by the more computationally expensive Delaunay tessellation
(see e.g. Okabe, Boots and Sugihara 1992). Donev, Torquato and Stillinger (2005) demonstrated advantage of the near-neighbor list method with respect to a traditional cell method (see e.g. Allen and Tildesley, 1987) used for speeding up {\color{black} for the}  neighbor search.
One wants to have the testing window $w_{\rm t}^{(km)}$ as large as possible so
that there is more room for the particle $v_i\subset v^{(k)}$ to move without leaving of $w_{\rm t}^{(km)}$. However, for the larger
$w_{\rm t}^{(km)}$, the more neighbors there will be to examine. The optimal balance, as determined by the
choice of $w_{\rm t}^{(km)}$ ($k,m=1,2$), is chosen numerically at the testing of concrete packs.
 Introduction of the testing window ${\color{black}w}_{\rm t}^{(km)}$ allows one to improve an efficiency of the shaking simulation if the displacement $\bfx_i'-\bfx_i$ is defined by possible collision of $v_i\subset v^{(k)}$ with $v_j\subset w_{\rm t}^{(km)}$. Namely, let $\bfx_i'-\bfx_i=\bfu_i^l\Delta t^l$ ($|\bfv_i^l|=1$) be a displacement of $\bfx_i$ in $l'$s trying of $\bfx_i$ random shaking where $\Delta t^l=\max _j\Delta t^l_j$, $v_j\subset w_{\rm t}^{(km)}$ and $\Delta t^l_j$ is obtained from a particular case of Eq. (8) ($j\in {\cal N}(i);\ l=1,\ldots, N^{(k)}_{\rm sh}$)
 \BB
 |\bfx_i+{\color{black}\bfu_i^l}\Delta t^l_j-\bfx_j|=a^{(k)}+a^{(m)}, 
 \EE
where the smaller positive root $\Delta t^l_j$ of Eq. (9) defines the collision time (or, by other words, the amplitude of $l'$s shaking) of inclusions $v_i$ and $v_j$. Found $\Delta t^l$ with the corresponding moving directions $\bfu_i^l$ allows one to find both the random amplitude of shaking uniformly distributed on the interval $[0,\tau]$ ($\tau=\min(\Delta t,R^{(k)}_{\rm sh})$, $\Delta t=\max_l\Delta t^l$) with the corresponding direction $\bfu^l_i$ of $v_i'$s moving.
Comparison of $\Delta t>R^{(k)}_{\rm sh}$ (or $<$) makes it possible to increase (or decrease, respectively) the size $R_{\rm sh}^{(k)}$ of the shaking window.

 Only the near neighbor set of inclusions
$v_j\ (j=j^i_1,\ldots,j^i_{n_{\rm t}}$; $n_{\rm t}$ depends on $k,m$) which are located in the testing window $w_{\rm t}^{(km)}$ centered at the point $\bfx_i$ ($v_i\subset
 v^{(k)},\ v_j\subset v^{(m)},\ k,m=1,2$)
are checked for overlap, which also reduces the computer time.
One determines the optimal size for the shake up window $R_{\rm sh}^{(k)}a^{(k)}$
 which provides the minimum average number of trial shaking attempts.
 This size of shake up window provides
the fastest stabilization of statistical parameter estimations.
However, the values of these parameters do not depend on the size of the shaking window
and are, furthermore, protocol independent (in contrast with the known methods
of random close packing simulations, see e.g. Torquato, Truskett and Debenetti, 2000).
The shaking process passes through all inclusion
(so called global shaking) with re-estimation of
the neighbor inclusions in the testing window $w_{\rm t}^{(km)}$
after each local shaking of the inclusion $v_i$.
In so doing, the increasing number of shaking has a twofold effect.
On the one hand, the system {\color{black}becomes} more homogeneous and well-mixed
and on the other hand, the stochastic fluctuations of statistical parameters (such as e.g. $g_{km}(r)$) estimated by averaging of these structures
tend to diminish.

The number of points in a sphere of radius $r$
about a randomly chosen point $\bfx_i$ is a random parameter. Estimation of its expectation defining $K$-function leads {\color{black}to loss} of statistical information
which can be presented in more details if one {\color{black}uses} a histogram form.
Namely, we evaluate the histograms of the
average number $n_{\rm h}$ of inclusions
in the spherical histogram window $w_{\rm h}^k=\{\bfx:\ |\bfx-\bfx_i|<R_{\rm h}^{(k)}a^{(k)}\}$
(i.e. the fractions $p(n_{\rm t})$ of the histogram windows containing the numbers $n_{\rm h}$ of inclusions) as a measure of inhomogeneity of an inclusion distribution inside
$w_{\rm h}^k$. Obviously, the descriptor $p(n_{\rm h})$ is a more sensitive measure of the local
statistical homogeneity of the configuration analyzed than the $K(r)$ is.

Thus, the scheme of the modified CRM (called shaking collective rearrangement model, SCRM) accompanied by the shaking procedure is the following.
After a few repetitions estimating
new velocities for the colliding growing inclusions (described in Subsection 3.1),
the inclusion shaking (described in Subsection 3.2) {\color{black}is} carried out. This new algorithm has a few advantages. The SCRM is not as optimal
as the original CRM for modeling close-packing configurations
simply because the added procedure of random shaking
is just focused on the ``destruction" of dense ``locked" local configurations
in some testing windows that would be manifested from one side by reduction of the first pick of the RDF $g(2a^+)$.
From other side, shaking leads to the generation of highly
homogeneous and mixed structures that can be controlled by the histograms of fractions $p(n_{\rm t})$ of testing window containing $n_{\rm h}$ particles.
Moreover, a distinguishing feature of the initial CRM is a large {\color{black}noise} of the estimated RDF (at least for the large particles, see numerical results in Section 5) because these data should be considered as merely a single realization of
such a random generation process. In order to provide statistically more reliable
results it would be necessary either to average several realizations or increase a size of a sample. However, this repeating
procedure is not necessary in the protocols accompanied by a shaking procedure
because a configuration generated by a few global shaking
can be regarded as a separate realization.

It should be mentioned that the shaking procedure is a very time consuming one. As it was mentioned, the shaking has twofold effect. On one hand side, sequential arrangement of global shaking destructs the dense locked local configuration generated by the CRM which is protocol dependent. Optimization of this part of algorithm from the point of view of CPU time reduction is hampered. However, because of the ideally parallel nature of the problem,
the second goal of shaking related with diminishing of stochastic fluctuations of statistical parameters estimated can be very
effectively reached by using of parallel computing at the standard
instruction-level parallelism that is performed in the computer codes developed.

\bigskip
\noindent {\bf 5 Numerical results}
\medskip

\noindent{\textit{\textbf{ 5.1 Unimodal distribution}}
\medskip

It should be mentioned that the most experimentally investigated cases of the RDF correspond to the close packing (from the loose packs $c=0.59$ to dense packs $c=0.64$), see
van Blaaderen and Wiltzius (1995), Gidaspov and Huilin (1998), Aste {\it et al.,} (2004, 2005), Saadatfar (2009),
Kurita and Weeks (2010). The volume fractions $0<c<0.6$ are less well understood, and, moreover, their {\color{black}investigations} are usually limited by suspensions with the different effects of interparticle interactions
(see, e.g., Sierou and Brady, 2002) that makes difficult their application to the solid composites. Estimation of the RDF from numerical simulations of random structures with $0<c<0.6$ is still less explored (see the use of the hard core model for simulation in Segurado and Llorca, 2002).

 The RDF $g(r)$ is well
investigated only for identical spherical ($3D$ and $2D$ cases, see for references, e.g., Buryachenko, 2007)
inclusions with a radius $a$. Two
alternative analytical RDFs of inclusion will be examined for 3D case
\BBEQ
g(r)\!\!\!\!&=&\!\!\!\!H(r-2a)\\ 
g(r)\!\!\!\!&=&\!\!\!\!H(r-2a)\left[{1+
g_c^{\small W}(c)g_r^{\small W}(r)
}\right], 
\EEEQ
where $H$ denotes the Heaviside step function, $r\equiv \vert
{\bf x}_i-{\bf x}_q\vert $ is the
distance between the nonintersecting fixed inclusions $v_i$ and moving one $v_q$,
 $c$ is the volume fraction of particles of the radius $a$; $g_r^{\small W}(r)=1$ at $r=2a^+$ and $g_c^{\small W}(c)\equiv g(2a^+)-1$.
The formula (10) describes a well-stirred approximation (differing from the RDF for a Poisson distribution by the availability of
``excluded volume" with the center $\bfx_i$ where $g(|\bfx_i-\bfx_q|)\equiv 0$)
while Eq. (11), proposed by Willis (1978), takes into account a neighboring order in the distribution of the inclusions (see also Stell and Rirvold, 1987):
\BB
g_c^{\small W}(c)=g_{c0}^{\small W}(c)\equiv 0.8006 \Big({8c-1\over c}\Big),\ \ \
g_r^{\small W}(r)=\cos \big({\pi r\over a}\big)e^{2(2-r/a)}. 
\EE
The restriction is that $c\geqslant 1/8$, when $g_{c0}^{\small W}(c)\geqslant 0$ ({\color{black}12}$_1$).
For avoiding of this restriction, Kanaun (1990) proposed a modification of the formula ({\color{black}12}$_1$) when $g_c^{\small W}(c)+1$ coincide with the first
peak of the popular Percus-Yervik (1958) approximation (see Throop and Bearman, 1965)
\BB
g_c^{\small W}(c)=g_c^{\small PY}(c)\equiv {2+c\over 2(1-c^2)}-1, 
\EE
 which is known to be most accurate when the density is not too high ($c<0.54$, see, e.g., Schaertl and Sillescu, 1994). Following Kanaun (1990), we will call Eq. (11) with the representations
 $g_c^{\small W}(c)$ (13) and $g_r^{\small W}(r)$ ({\color{black}12}$_2$) as the Willis approximation (11).

The analytical RDFs (10), (11) and Percus-Yervik (P-Y, 1958) approximation (see numerical results for P-Y approximation in Throop and Bearman, 1965) are compared with the numerical simulations by the CRM and SCRM (the number of spheres {\color{black}is} $n=2000$) for $c=0.575$ in Figs. 1 and 2 for
the ranges of a normalized coordinates $2<R\equiv r/a<2.2$ and $2.5<R<4.5$, respectively. As it is well

\vspace{-20mm}
\noindent \parbox{7.8cm}{
\hspace{-20mm} \psfig{figure=figI_1.ps, width=12.0cm}\\
\vspace{-80mm} \tenrm \baselineskip=8pt

{\rm Fig. 1}:
The RDFs $g(r)$ vs $r/a$ at $c=0.575$ estimated by
CRM (1), SCRM (2), Eq. (5.2) (3), P--Y (4)
} \ \ \noindent \parbox{7.8cm}{\vspace{0mm}
\hspace{-30mm}
\psfig{figure=figI_2.ps,width=12.0cm}\\
\vspace{-80mm}\tenrm \baselineskip=8pt

{\rm Fig. 2}. The RDFs $g(r)$ vs $r/a$ at $c=0.575$ estimated by
CRM (1), SCRM (2), Eq. (5.2) (3), P--Y (4) }

\vspace{2mm}

\noindent known, the values $g_c^{\small PY}(c)=g_c^{\small W}(c)$ are underestimations
of the corresponding first picks at $r=2a^+$ obtained by both the CRM and SCRM.
It should be mentioned that the Monte Carlo shaking process is both the most structurally influential and,
computationally, the most  intensive part of the whole packing process.
So, packing by the CRM for $n=2000$ takes a few seconds for a PC with a 2.4 GHz processor while one global shaking requires 5 min and stabilization
(rather than elimination of a statistical {\color{black}noise} of descriptors evaluated) of statistical parameter estimations is reached after 10 global shakings.
Thus, the shaking process is very time-consuming and sensitive to a few competitive variables (such as $R^{(1)}_{\rm sh}$ and $N^{(1)}_{\rm sh}$) and
 the frequency ratio of the alternation of the CRM's iterations and the shaking
which should be optimized (optimization of the CRM algorithm
was considered in Maggi, {\it et al.}, 2008). So, increasing of
the shaking radius $R^{(1)}_{\rm sh}$ leads to both the positive and negative effects. From one side, increasing of $R^{(1)}_{\rm sh}$ leads to large amplitude of possible moving of an inclusion $v_i$ but from other side, it {\color{black}brings} necessity to increase both the
  near-neighbor list ${\cal N}(i)$
 and the limiting shaking number $N^{(1)}_{\rm sh}$ that is very time-consuming. The optimal parameters $R^{(1)}_{\rm sh}=0.02$ and
 $N^{(1)}_{\rm sh}=150$ were found at the data estimations in Fig. 1 and 2. It should be mentioned that derivative in the RDF (4) is estimated by the method of finite difference with a chosen $\Delta r$. If $\Delta r$ is too large, $g(r)$ is smoothed and peaks are obscure; if too small $g(r)$ has large statistical {\color{black}noise} with perhaps spurious peaks
 because only a small number of particles can be observed in a spherical shell with the thickness $\Delta r$ and
 an inner radius $r$. Some trial runs indicate the optimal $\Delta r=0.01a$; decreasing $\Delta r$ from $0.015a$ to $0.01a$ leads to increasing of the first peak at $r=2a^+$ on $3\%$ while the different choices of $\Delta r$ within a broad range of $10^{-4}a$ to $10^{-2}a$ lead to almost indistinguishable results.

 {\color{black} It is interesting that the estimations of $g(r)$ (4) can be performed  through the finite difference analogs of the central (Aste {\it et al.}, 2004) and right
(Maggi, Stafford, and Jackson, 2009) derivatives ($\Delta r>0$)

\BBEQ
g^{\pm}(r,\Delta r)\!\!\!&=&\!\!\!
b[K(r+\Delta r/2)-K(r-\Delta r/2)](\Delta r)^{-1},\ \ \ \nonumber \\
\ g^{+}(r,\Delta r)\!\!\!&=&\!\!\! b[K(r+\Delta r)-K(r)](\Delta r)^{-1},  
\EEEQ
respectively, where the constant $b=r^{1-d}/(d\omega_d)$ is fixed by imposing that asymptotically $g(r)\to 1$ as $r\to \infty$. The formulae (14$_1$) and (14$_2$) providing the accuracies
$O(\Delta r^2)$ and $O(\Delta r)$, respectively, lead to close results at $\Delta <10^{-2}a$ if
$K(r)$ is differentiable in $r$. However, $K(r)$ has discontinuity at $r=2a$ and is not differentiable at $r=2a$: $K(r-0)=0$ and $K(r+0)>0$. Therefore, a formal incorrect use of Eq. (14$_1$) leads to the dramatic result at $r=2a$:
\BB
g^{\pm}(r,\Delta r)=b{K(r+\Delta r/2)\over \Delta r} ={1\over 2} g^+(r,\Delta r/2), 
\EE
and, therefore at $r=2a$
\BB
g^{\pm}(r)=g^+(r)/2  
\EE
while
$g{\pm}(r)=g^+(r)$ at $r>2a$; here $g^{\pm}(r)=\lim g^{\pm}(r,\Delta r)$
and $g^{+}(r)=\lim g^{+}(r,\Delta r)$ for $\Delta r\to 0$. Apparently the use of Eq. (16) has led
Aste {\it et al.} (2004) to an unusual power law approximation $g(r)\propto(r-r_0)^{-\alpha}$ in the range $r\leq 2.8a$ with the singularity at $r_0=2.06a$ while a similar behavior, but with $r_0=2a$
(that agree with our results), was reported by Silbert {\it et al.} (2002) for molecular dynamic simulations.
}

The RDFs estimated from the CRM and SCRM simulations in Figs. 1 and 2 have a well-known form
(see, e.g., Lochmann, Oger and Stoyan, 2006; Maggi {\it et al.}, 2008) with the main maxima appearing at well-defined positions,
namely at $r=2a, 2\sqrt{3}a, 4a$ ({\color{black} the first, second, and third peaks, respectively  )} and so on (which can be also presented in a nondimensional form $R=2,\ 2\sqrt{3},\ 4$) and then $g(r)$ continues to fluctuate with decreasing amplitude. The smoothing of the RDFs obtained from the SCRM simulation was performed by the averaging over $P=50$ realizations of paralleled simulations.
These maxima represent the distances between the centers of a
reference sphere to its first, second, third neighbors etc.
The first very pronounced peak (discontinuity) at the hard-core distance $r=2a$ is a result of a direct contact of spheres which is a most frequent nearest-neighbor distance.
The second peak at $r=2\sqrt{3}a\approx3.464a$ (see Fig. 2, and 3(A)) corresponding to a {\color{black}Face-Centered Cubic} (FCC) lattice and appearing at the high packing fraction is completely lost at the {\color{black}low volume fraction of particles}. The peak at $r=4a$ (growing faster with growing $c$ {\color{black}than} the second one) sorts well with the placement of  three {\color{black}touching} spheres in a row (see Fig. 3(B)). The
 gap distance $r=2.75a\doteqdot2.85a$ corresponding to the empty  space between the first and second neighbors of typical particle is also clearly observable.
The third {\color{black}peak} at $r=4a$ is well predicted by the Percus-Yervik (1959) approximation which does not expect the second {\color{black}peak} at
$r=2\sqrt{3}a$ for any $c$. Willis (1978) approximation (11) (curves 3 in Figs. 1 and 2) is in agreement with simulation observation (curves 1 and 2) only qualitatively.
Arrangement of inclusions presenting in Figs. 3(A) and 3(B) the fixed left inclusion and defining the second and third, respectively, peaks can be denoted in the
forms $v_i-(v_i+v_i)-v_i$ and $v_i-(v_i)-v_i$, respectively. Here the inclusions placed within the parentheses represent the first shell of nearest moving particles.

 As can be seen in Fig. 2, the SCRM leads to the smoothing of the RDFs $g^{\small SCRM}(r,c)$.
 But the smoothing of the RDFs (see {\color{black}the} curve 1 in Fig. 2) could be reached significantly cheaper in the framework of the initial
version of the CRM by increasing of either $N$ or the number of realizations $X$.
 Nevertheless, any smoothed RDF obtained by averaging of RDFs of configurations generated by the CRM don't remove the birth-marks of a fundamental shortcoming of the CRM such as protocol dependent and too strong local arrangement in the nearest zone ($r=2a^+$).

 In doing so, a fundamental advantage of the SCRM is, first of all, a destruction of
 dense ``locked" local configurations (which are protocol dependent) leading to the most homogeneous and mixed structures. It is especially pronounced in Fig. 1 where SCRM demonstrates a reduction of $g(2a^+)$ estimated by the CRM {\color{black}of} 43\% (destruction of the nearest zone arrangement which is responsible for the local jamming). For more detailed analysis of RDF's regularities, one estimates $g(r)$ by both the CRM and SCRM (at the same simulation parameters as in Figs. 1 and 2) for more dense packing at $c=0.6$ and $c=0.63$. As it is expected, the CRM's peaks are always greater than the SCRM's ones, significantly so in some cases. For example the SCRM reduces the first peaks ($r=2a^+$) estimated by the CRM {\color{black}of} 43\%, 21\%, and just 4\% at $c=0.575,\ 0.6$, and $0.63$, respectively. In doing so, the shaking procedure at $r>2.5a$ leads to nothing more than a smoothing at any volume fractions $c$ of inclusions; so, even for $c=0.6$ a systematic reduction of the {\color{black} peaks} at $r=2\sqrt{3}a$ and $r=4a$ is no better than $10\%$ while there {\color{black}is} no systematic difference outside of peak's areas at $r>2.5a$ between RDFs of configurations simulated by both the CRM and SCRM.
 So {\color{black}much} significant reduction of $g(2a^+)$ at $c=0.575$ and $c=0.60$, is explained by the fact that at the volume fractions $c$ far from the jamming limit $c=c^{\rm jam}=0.6366$ a particle configuration has many local areas with relatively small densities where the particles can be removed from the {\color{black}destructed nearest} zones $r=2a^+$ by the shaking.
 The last statements is confirmed by {\color{black}Fig. 4} where the histogram of fractions $p(n_{\rm h})$ of histogram window $r_{\rm h}=2.5a$ containing $n_{\rm h}$ inclusions
 are presented for $c=0.6$. As can be seen, shaking leads to reduction of the numbers of histogram windows

\vspace{-20mm}
\noindent \parbox{7.8cm}{
\hspace{-20mm} \psfig{figure=figI_3.ps, width=12.0cm}\\
\vspace{-80mm} \tenrm \baselineskip=8pt

{\rm Fig. 3}:
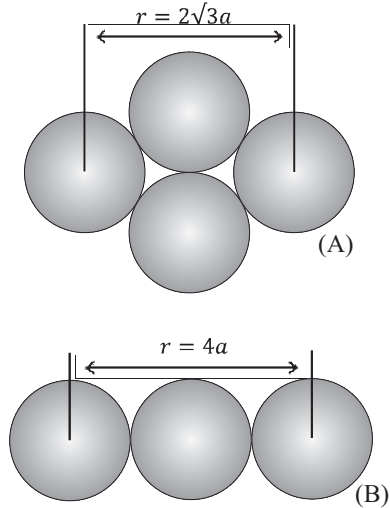
Possible sphere configurations and the peak locations in the RDF: (A) Four close spheres
$v_i-(v_i-v_i)-v_i$, (B) Tree spheres in a row $v_i-(v_i)-v_i$
} \ \ \noindent \parbox{7.8cm}{\vspace{0.mm}
\hspace{-20mm}
\psfig{figure=figI_4.ps,width=12.0cm}\\
\vspace{-80mm}\tenrm \baselineskip=8pt

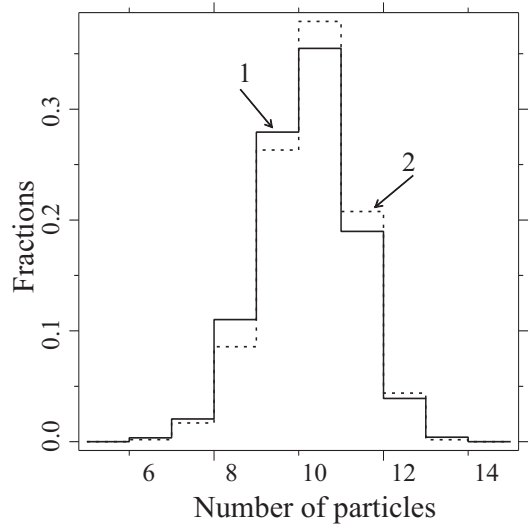
{\rm Fig. 4}. Histogram of fractions $p(n_h)$ of histogram window containing nh spheres
generated by the CRM (solid line, 1), SCRM (dotted line, 2)
}
\vspace{2mm}


 \noindent with small and largest numbers of particles and increasing of window number with average number of spheres.
 However, at the $c\geq0.63$ the areas with local dense ``locked" packing are increasing in the volume and occupy in the limiting case $c=0.6366$ a full sample. Because of this, shaking has very poor effectiveness at $c=0.63$ and reduces the first peak at $r=2a^+$ just {\color{black}by} 4\%.

The availability of 3D imaging tools such as computed X-Rays tomography and microtomography, confocal optical microscopy or magnetic resonance imaging, combined with the increased power of computers allows us to process large quantities of data (including the coordinates of particle centers) at reasonable costs.
It is important to be {\color{black}sure} that the packs simulated by proposed virtual methods have the same statistics as the
real material packs. It is possible, of course, that real and simulated packs will have
different statistics according to the manner in which they are
manufactured and generated, a matter that we are not in a position to address at
the present time. But Aste and his colleagues at the Australian
National University in Canberra have measured the RDFs for
a number of very large monodisperse packs ($\sim150,000$ spheres) using X-ray tomography
(Aste {\it et al.}, 2004, 2005), and have provided us with their raw data so that we
might make comparisons, and we do that here. Our packs use
35,000 spheres in a
 periodic cube and are generated using the
T algorithm {\color{black}(see for details Maggi {\it et al.}, 2008)} with packing fractions of 59.25\%; the experimental packing fractions are 59.3\%. The comparisons are shown in Figs. 5 and 6. As can be seen, the curves simulated by the SCRM ($n=2000;\ P=80$) is better


\vspace{-20mm}
\noindent \parbox{7.8cm}{
\hspace{-25mm} \psfig{figure=figI_5.ps, width=12.0cm}\\
\vspace{-80mm} \tenrm \baselineskip=8pt

{\rm Fig. 5}:
The RDFs $g(r)$ vs $r/a$ at $c=0.593$ estimated by
CRM (1), SCRM (2), Aste's data (3)
} \ \ \noindent \parbox{7.8cm}{\vspace{0mm}
\hspace{-25mm}
\psfig{figure=figI_6.ps,width=12.0cm}\\
\vspace{-80mm}\tenrm \baselineskip=8pt

{\rm Fig. 6}. The RDFs $g(r)$ vs $r/a$ at $c=0.593$ estimated by
CRM (1), SCRM (2), Aste's data (3) }

\vspace{2mm}

\noindent  correlated with experimental data by
Aste {\it et al.} (2004, 2005) (see also Silbert {\it et al.}, 2002) than CRM's simulation. In doing so, there is no reason for expectation that RDFs estimated from experimental data are the unbiased ones which can be correctly compared with the RDFs evaluated from the simulated data.
Indeed, both the CRM's and SCRM's data were simulated for unimodal distribution with identical sphere size $a$ while for real polydisperse, non-perfect, spheroidal particles, the distance between the centers of particles in contact is not a fixed value $2a$ but instead it is distributed around an average value. However, exploiting of Eqs. (2) and (4) implies the use of some fixed radius $a_{\rm f}$ of particles which was estimated from experimentally found volume fraction $c$ and the intensity $n$ according to the formula $4\pi a_{\rm f}^3n/3=c$, or, by other words, $a_{\rm f}=\lle a^3(\zeta)\rle^{1/3}$, where $a(\zeta)$ is a random radius of particles. This defines a bias of the RDF evaluated from experimental data. So, the use of $a_{\rm f}=1.05\lle a^3(\zeta)\rle^{1/3}$ and $a_{\rm f}=0.95\lle a^3(\zeta)\rle^{1/3}$ instead of $a_{\rm f}=\lle a^3(\zeta)\rle^{1/3}$ leads to the variation of the first peak $g(2a^{+})$ on 40\% (reduction) and 7\% (increasing), respectively.


 It should be mentioned that for the SCRM's configurations, the particle reorganization induced
by shaking is subjected only to geometrical constraints, whereas for real structures the packing is far more complicated and controlled by elastic, hydrodynamic, cohesive forces, and chemical specificity etc. (see, e.g., Alcaraz {\it et al.}, 2008). Our simulation technique is
able to isolate the fundamental geometrical constraints from other physio-mechanical and chemical effects
and therefore the results provide a valuable benchmark for evaluating sophisticated packing schemes used
in the modeling of real composite materials. It is shown that a sufficiently intensive shaking process leads
to the stabilization of a statistical distribution of the simulated structure that is most homogeneous, highly mixed
and protocol independent (in sense that the statistical parameters estimated do not depend on either the
basic simulated algorithm such as, e.g., CRM or manufacturing parameters of the real material production). It is expected (and proved for 2D case by Buryachenko {\it et al.}, 2003) that
shaking procedure will reduce the peaks (having a clustered nature) of experimentally estimated RDF's (by the use of, e.g., X-ray tomography, see Aste {\it et al.}, 2004, 2005) rather than only the peaks of RDFs generated by the CRM (the arguments justified the last statement are
plausible rather than rigorous and require additional investigations). This inconsistency between the real and SCRM's structures
is not a deficiency of the SCRM because a subsequent use of any advanced micromechanical model (see for references Buryachenko, 2007) assumes a stationary of a process $X$ rather than its slightly clustered configuration as for real materials.

 Due to a costliness of the SCRM, it can be useful to extrapolate a value $g^{\small SCRM}(r)$ estimated at the volume fraction $c_1$ to other particle concentration $c_2$. It can be easy done if one assumes that $g^{\small SCRM}(r)$ has in vicinity $c=c_1$ the same functional dependence on $c$ and $r$ as Willis's approximation (12): $g^{\small SCRM}(r,c)=H(r-2a)[1+g^{\small SCRM}_c(c)\cdot g^{\small SCRM}_r(r)]$ but with the different
$g^{\small SCRM}_c(c)$ and $g^{\small SCRM}_r(r)$. Assuming also that $g^{\small SCRM}_c(c_1)/g^{\small SCRM}_c(c_2)=g^{\small W}_c(c_1)/g^{\small W}_c(c_2)$, we can obtain
 \BB
 g^{\small SCRM}(r,c_2)-1= [g^{\small SCRM}(r,c_1)-1]\cdot [g^{\small W}_c(c_2)-1]/ [g^{\small W}_c(c_1)-1]. 
 \EE

It should be mentioned that the RDF was only defined by Eq. (4) for statistically homogeneous and isotropic fields of inclusion centers $\bfx_1,\bfx_2,\ldots$.
Then $f_2(\bfx-\bfy)$ and $g(r)$ are the continuous isotropic functions. However, for {\color{black}the} lattice structures,  $f_2(\bfx-\bfy)$ is a sum of a delta functions (with the support at the particle centers) and correctness of using of Eq. (4)  is not obvious because in such a case estimators of $g(r)$ for a stationary ergodic process are difficult to interpret (see Picka, 2007). {\color{black} However, due to the practical importance of lattice structure investigations, we present some correct interpretations of Eq. (4) for these cases. Namely, we consider as an example a simple cubic lattice ($t=x,y,z$): $t_i=i\times 2a$ ($i=0,\pm 1, \pm 2,\ldots,  \pm n_l$, $2a=1/(n_l+0.5)$.  At first we estimated $g^+ (r,\Delta r)$ (14$_2$) for $n_l=1, 2, 5$ and the normalized radius $R=r/a$ and the radius steps $\Delta R=\Delta r/a=0.1, 0.01, 0.005, 0.003$: we get a picture similar to Nolan and Kavanagh (1995) with $ g^+ (2,\Delta R)=9.55, 95.5, 190.9$, and $318$, respectively, i.e. $g^+ (2,\Delta R)\to \infty$ for $\Delta R\to 0$ that is intrinsic for the delta function, e.g. $\delta (r-2a)=d H(r-2a)/dr$.  However, any integral over the domain containing $R=2$ eliminates the delta function that gives different opportunities for interpretation of  $g^+ (r)$. In particular, the integral over the shell with the internal radius $r$ and the thickness $\Delta r$
\BB
I(r,\Delta r)=4\pi n\int^{r+\Delta r}_r r^2  g^+ (r,\Delta r)dr=[4\pi n r^2 \Delta r]×[g^+ (r,\Delta r)], 
\EE
(where $n=N/\overline w$) is a number of inclusion centers in the shell.  It was detected that for any $n_l$ (at least for $n_l=1, 2, 5$) and any $\Delta R$ (at least for $\Delta R=0.1, 0.01, 0.005, 0.003$), the integers $I(r,\Delta r) = I(r)$ does not depend on $n_l$ and $\Delta r$. Thus, both items in the square brackets in Eq. (18) depend on  $\Delta r$  but its product $I(r,\Delta r)$  does not depend on $\Delta r$ (at least at $\Delta r<0.1$).
A plot $I(r)$ vs $r/a$ is presented in Fig. 7. A visually continuous curve $I(r)$ vs $r/a$ (instead of discontinuous integer function in analytical solution) is obtained due to numerical discretization of the delta-function. As we can see, the first peak coincides with the number of the nearest neighbors in the simple cubic packing being considered. Thus, the integral $I(r)$ uniquely defined  has an obvious physical meaning (in contrast to $g(r)$).
}

\noindent \hspace{22mm}
\parbox{7.8cm}{\vspace{0.mm}
\hspace{20mm}
\psfig{figure=figI_7.ps,width=7.2cm}\\
{\vspace{-0mm}\hspace{10mm}\rm \baselineskip=8pt

\hspace{10mm}{\rm Fig. 7}. $I(r)$ vs $r/a$ for a simple cubic lattice}}


\bigskip \noindent
\noindent{\textit{\textbf{ 5.3 Bimodal distribution}}
\medskip

For polydisperse structures the counterpart to the RDF is less investigated than for monodisperse case and described by the partial RDF $g_{ij}(r)$ which is the normalized (by the condition $g_{ij}(\infty)=1$) number density of inclusions $v^{(j)}$ of type $j$ that are within shells $r$ to $r+\Delta r$ centered on the inclusions $v^{(i)}$ of type $i$. $g_{ij}(r)$ were estimated by the modified CRM by Maggi {\it et al.} (2008) for the size ratio $\lambda$ of spherical inclusions more {\color{black}than} $6.5$. As it was expected, for the large inclusions, $g_{11}(r)$ is stochastically very noisy, although trends are apparent. Because of this, as we have already noted, incorporation of the shaking procedure in the proposed algorithm is very effective for smoothing of $g_{ij}(r)$.

It is clear that we have to measure the distances between the particles $v_i\subset v^{(k)}$ ($a_i=a^{(k)}$) and $v_j\subset v^{(m)}$
($a_j=a^{(m)}$) as the excess distances with respect to contact, that is $\Delta r^{\rm con}=r-(a_i+a_j)$; rather than the distances $r$ themselves. Then the term $H(r-2a)$ in Eqs. (10) and (11) should be replaced by $H(r-a_i-a_j)$
Then, we have
essentially two ways of scaling these excess distances. One is to take the same scaling
parameter for the distances in all partial RDFs $g_{km}$ ($k,m=1,2)$ of the mixture. This is the
situation in the decoupling approximation (see
 Kotlarchyk and Chen, 1983;  Barrio and Solana, 2005;
see also Lochmann, Oger and Stoyan, 2006)
in which the scaling parameter is
unity, but we can take equally {\color{black}use} other suitable scaling parameter such as $a^{(m)}$.
In such a case a different scaling distance of each partial RDF for the second peak
generated by the second shell of moving particles $v_i\subset v^{(m)}$ are placed approximately at the same positions, although the height of the peaks corresponding to different RDFs are different. This is particularly remarkable for the first peak at $r=a_i^++a_j^+$. Thus, we will present the partial RDFs $g_{km}$ as the functions of a scaling parameter $R_{km}=(r-a^{(k)})/a^{(m)}+1$. In a general case, the peaks are a consequence of the {\color{black}bidisperse} morphology, but some of them can be explained within the monodisperse framework.
For example, the coordinates of the first peak of
$g_{km}$ and $g_{mk}$ for $k \not = m$ coincide in both arguments $R_{km}=2$ and $r=a^{(k)}+a^{(m)}$ while the second and third peaks depends on additional information. {\color{black}The} nature of the third peak at $R_{km}=4$ (or $r=a^{(k)}+3a^{(m)}$) described within the monodisperse framework reflects the placement of the touching inclusions $v_i-(v_j)-v_j$ ($v_i\subset v^{(k)},\ v_j\subset v^{(m)})$ in a raw (see Fig. 3B). However, a specific bidisperse morphology can be manifested
by a lot of additional specific configurations.
So, it can be generated by the configuration $v_i-(v_i)-v_j$ of touching inclusions in a raw that produces a new peak at $r=3a^{(k)}+a^{(m)}$ or, in normalized coordinates, at $R_{km}=2(1+\lambda_{km})$ ($\lambda_{km}=a^{(k)}/a^{(m)}$). Analogously, the related peaks of $g_{mk}(r)$ corresponding to the configurations
$v_j-(v_i)-v_i$ and $v_j-(v_j)-v_i$ are placed at $r=a^{(k)}+3a^{(m)}$ and $r=3a^{(m)}+a^{(k)}$, respectively, that matches with the peak coordinates of $g_{km}(r)$.
Thus, the third peak ($R=4$) of unimodal pack has two possible counterparts of their bidisperse pack with
$R_{km}=4$ and $R_{km}=2(1+\lambda_{km})$ coinciding only at $\lambda_{km}=1$. In a similar manner, the second peak of unimodal pack with a symbolic configuration $v_i-(v_i+v_i)-v_i$ (see Fig. 3A) have three counterparts for the fixed inclusion $v_i\subset v^{(k)}$:
$v_i-(v_i+v_i)-v_i$, $v_i-(v_j+v_j)-v_j$, and $v_i-(v_i+v_j)-v_j$. For example, the configurations $v_i-(v_i+v_i)-v_j$ and $v_i-(v_j+v_j)-v_j$ {\color{black}corresponding} to the peaks $R_{km}=(\sqrt{3}-1)\lambda_{km}+\sqrt{1+2\lambda_{km}}+1$ and $R_{km}=\sqrt{3}-\lambda_{km}+\sqrt{\lambda_{km}^2+2\lambda_{km}}+1$, respectively.
Of course {\color{black}for other} the volume fractions $c$, size ratio $\lambda_{km}$, and volume fraction ratio $\xi_{km}=c^{(k)}/c^{(m)}$ ($k\not =m$) being considered, one obtains different results for the peak locations (see, e.g., Lochmann, Oger and Stoyan, 2006) when the different ``monodisperse" and ``bidisperse" peaks are pronounced in varying degrees. So, visualization of peaks in unimodal packs is only managed by a single parameter $c$ while the peaks of bimodal packs are managed by three parameters $c,\lambda_{km}$, and $\xi_{km}$ ($k\not=m$).

Due to the absence of the analytical representations of $g_{ij}(r)$ ($i,j=1,2$) for polydisperse structures, it is possible to construct $g_{ij}(r)$ within monodisperse considered framework (10) and (11) in the following manner ($R_{ij}=(r-a_i)/a_j+1$):
\BBEQ
g_{ij}(r)\!\!\!\!&=&\!\!\!\!H(R_{ij}-2),\\ 
g_{ij}(r)\!\!\!\!&=&\!\!\!\!H(R_{ij}-2)\left[{1+ [g^c_{ij}(\xi,c)-1]
\cos ({\pi R_{ij}})e^{2(2-R_{ij})}}\right], 
\EEEQ
where $c=c^{(1)}+c^{(2)}.$ The RDFs contact values $g^c_{ij}(\xi,c)$ were obtained by Attard (1989)
from {\color{black}Ornstein-Zernike} equation using Percus-Yevick closure of the pair and triplet levels.  In particular, the P-Y result obtained for the bulk mixture at the pair level is
\BBEQ
g^c_{ij}(a_{ij})&=& [a_j g_{ii}(a_{ii})+a_i g_{jj}(a_{jj}]/a_{ij}, \\ 
g^c_{ii}(a_{ii})&=& [1-\zeta_3+3a_i\zeta_2](1-\zeta_3)^2, 
\EEEQ
where $a_{ij}=a_i+a_j$ is a mutual diameter, and $\zeta_{\alpha}=(\pi2^{\alpha-1}/3)\sum_ia_i^{\alpha}$. Taking the limit as $c^{(2)}\to 0$, so that $\zeta_3=c^{(1)}$, one has
\BB
g_{12}^c(a_{12})=\Big[1+c^{(1)}{(2\lambda_{21}-1)\over (\lambda_{21}+1)}\Big]{1\over (1-c^{(1)})^2}, 
\EE
which reproduces the respective limit $g_{12}^c(a_{12})=(1+c^{(1)}/2)/(1-c^{(1)})^{-2}$ corresponding to the usual P-Y value for unimodal pack $g_c^{PY}(c)$ (20).

The paper by Chong, Christiansen and Baer (1971)
is dedicated to the experimental measurement of effective Newtonian suspensions $\eta^*/\eta^{(0)}$ of spherical bimodal particles with the size ratio $\lambda\equiv \lambda_{21}= a^{(2)}/a^{(1)}=0.477,0.313,$ $0.138$, and
the volume fraction ratio $\xi\equiv \xi_{21}=c^{(2)}/c^{(1)}=0.333$. Because of this one considers the partial RDF's $g_{ij}(R_{ij})$ (no summation over $i,j=1,2$) simulation for bimodal packs with $\lambda=0.313$ and $c=0.6$. The typical first peak
at $R_{ij|1}=2$ of the $g_{ij}(R_{ij})$ presented for the RDF $g_{11}(R_{11})$ in Fig. 8 is analogous to the first peak for the unimodal pack (see Fig. 1);
shaking {\color{black}parallel} procedure SCRM yields a reduction of the first CRM's peak on 25\%. The curves $g_{ij}(R_{ij})$ (see the curves 2 in Figs. 9-11) behave at $R_{ij}>2.2$ less trivial than $g(R)$ (see Fig. 2). So the second peak $R_{11|2}=2.64$ holds a
clearly defined bidisperse nature and corresponds to the configuration $v_1-(v_2)-v_1$: $R_{11|2}=2.64\simeq 1+2\lambda+1$. The ``monodisperse" peaks $R_{11|3}=3.464$ and $R_{11|4}=4.0$ correspond to the peaks $R=2\sqrt {3}\simeq 3.462 $ and $R=4$ of
the unimodal pack in Fig 2.
The pronounced peak in
Fig. 9 at $R_{11|5}=4.54$ corresponds to the case when the shell either $-(v_2+v_2)-$ or $-(v_2)-$ is
  wedged
in between the three spheres in a row $v_1-(v_1)-v_1$ that moves the
peak $R_{11|4}=4.0$
on either $2(\sqrt{1+2\lambda}-1)\simeq0.55$ (see Fig. 3A) or $2\lambda\simeq 0.62$ (see Fig. 3B), respectively, that is matched up to the average displacement $0.59$. It is interesting that the first gap distance $R=2.8\doteqdot 2.9$ in unimodal pack (see Fig. 1) corresponds to the empty space between the


\vspace{-20mm}
\noindent \parbox{7.8cm}{
\hspace{-20mm} \psfig{figure=figI_8.ps, width=12.0cm}\\
\vspace{-75.mm} \tenrm \baselineskip=8pt

{\rm Fig. 8}:
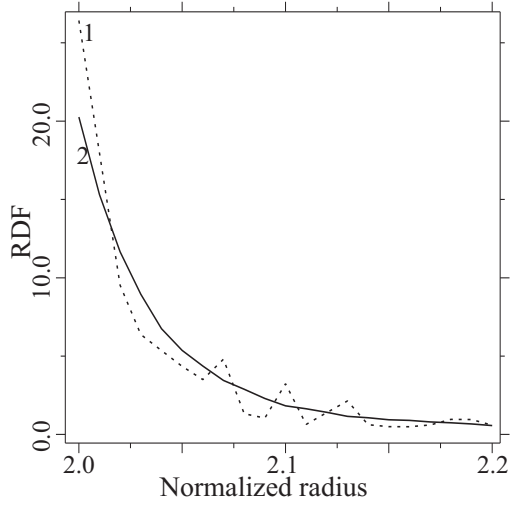
$E^*(c)/E^{(0)}$ vs $c$ estimated by the MEFM with RDF simulated by CRM (1), SCRM (2), Willis approximation (3), well-stirred approximation (4), and by the MTM (5); $\circ$ - experimental data by Smith (1976)}
 \ \ \noindent \parbox{7.8cm}{\vspace{1.mm}
\hspace{-20mm}
\psfig{figure=figI_9.ps,width=12.0cm}\\
\vspace{-80.mm}\tenrm \baselineskip=8pt

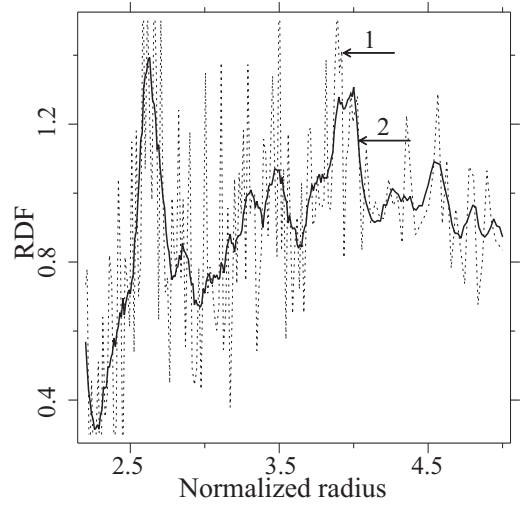
{\rm Fig. 9}. $\mu^*(c)/\mu^{(0)}$ vs $c$ estimated by the MEFM with RDF simulated by
Willis approximation (1), SCRM's approximation (5.5) (2), CRM (3), SCRM (4), PY (5), well-stirred approximation (6), and by the MTM (7); $\circ$ - experimental data by Kreger (1972)
}
\vspace{2mm}

\noindent first and the second neighbors of typical particle $v_1$ and symbolically {\color{black}denoted} by the form $v_1-(v_1+v_1)\smile$ at $R=2+\sqrt{3}/2\simeq 2.87$. This gap is manifested in $g(R)$ but not in $g_{12}(R_{12})$ and $g_{22}(R_{22})$
where the first gap distance is formed by a bidisperse configuration $v_1-(v_1)\smile$ and
$v_2-(v_1)\smile$ at $R_{12}=1/\lambda\simeq 3.2$
 and $R_{22}=1/\lambda\simeq 3.2$ (observed as 3.3 in Figs. 10 and 11), respectively,
rather than by a configuration $v_1-(v_2+v_2)\smile$ and
$v_2-(v_2+v_2)\smile$ at $R_{12},R_{22}=2.87$. The distance $R_{12},R_{22}=1/\lambda\simeq 3.2$ corresponds to the
middle surface of the shell $-(v_1)-$ surrounding a typical fixed point $v_1$ (or
$v_2$), occupied by the centers of moving particles $v_1$ and inaccessible to the centers $v_2$. Moreover, a similar gap at the distance $R_{11}=2+\lambda\simeq 2.3$ (see Fig. 9) of $g_{11}(R_{11})$ holds the same bidisperse nature and is produced by the shell $-(v_2)-$ surrounding a typical point $v_1$ and unattainable to the centers of moving particles $v_1$. The next (after $R_{12}=4$) most pronounced peak of $g_{12}(R_{12})$ in Fig. 10 is analogous to the peak $R_{12}=4$ with the configuration $v_1-(v_2)-v_2$ where a middle sphere $-(v_2)-$ is replaced by a sphere $-(v_1)-$ in the configuration $v_1-(v_1)-v_2$ that leads to displacement of the peak $R_{12}=4$ on $2/\lambda-2$ to the right to $R_{12}=2+2/\lambda\simeq 8.3$. In a similar {\color{black}manner}, the peak of $g_{22}(R_{22})$ in Fig. 11 at $R_{22} =2+2/\lambda\simeq 8.3$ produced by the configuration $v_2-(v_1)-v_2$ is clearly defined.

The RDFs curves {\color{black}$g_{11}(R_{11})$,} $g_{12}(R_{12})$ and $g_{22}(R_{22})$ in Figs. 8-11 obtained for $N=2000$ were smoothed by averaging over $P=80$ realizations of the paralleled SCRM's generations. For analysis of the impact of a {\color{black} sample size} $N$, one performs a similar modeling for $N=7000$ for $M=3$ global shaking and compares it with analogous simulations for $N=2000$ with $M=3$ and $M=60$ global shaking (see Figs. 12 and 13). As can be seen the RDFs
estimated for the samples with $N=2000,\ M=60$ and $N=7000,\ M=3$ are
close while the difference of the RDFs estimated for $N=2000$  with
$M=3$ and $M=60$ global shaking is significant. The results of comparison indicate on a dual nature of smoothing averaging which accuracy can be increased by
 two ways either by increasing of the number of realizations (the number of paralleled estimations $P$) or by a growing of a sample size (the number $N$). It means that a sample with $N=2000$ can be considered as a statistically representative of an infinite sample. A remarkable difference between the RDFs estimated for $N=2000$ with $M=3$ and $M=60$
global shaking points to the fact that $M=3$

\vspace{-17mm}
\noindent \parbox{7.8cm}{
\hspace{-20mm} \psfig{figure=figI_10.ps, width=12.0cm}\\
\vspace{-75mm} \tenrm \baselineskip=8pt

{\rm Fig. 10}:
The RDFs $g_{11}(r)$ vs $R_{11}$ at $c=0.6$ estimated by
CRM (1), SCRM (2)
} \ \ \noindent \parbox{7.8cm}{\vspace{0mm}
\hspace{-18mm}
\psfig{figure=figI_11.ps,width=12.0cm}\\
\vspace{-80mm}\tenrm \baselineskip=8pt

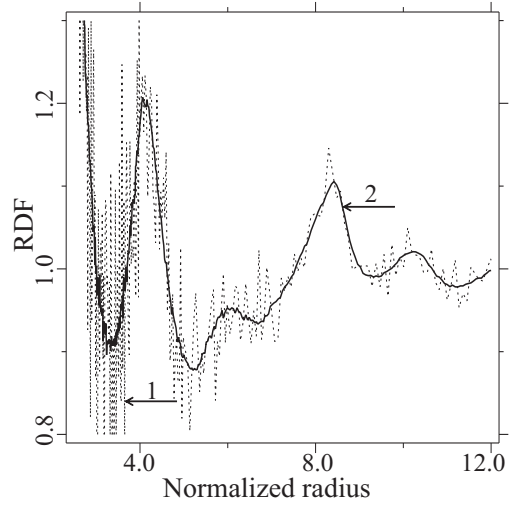
{\rm Fig. 11}. The RDFs $g_{11}(R_{11})$ vs $R_{11}$ at $c=0.6$ estimated by
CRM (1), SCRM (2) }
\vspace{2mm}

\vspace{-17mm}
\noindent \parbox{7.8cm}{
\hspace{-20mm} \psfig{figure=figI_12.ps, width=12.0cm}\\
\vspace{-80mm} \tenrm \baselineskip=8pt

{\rm Fig. 12}:
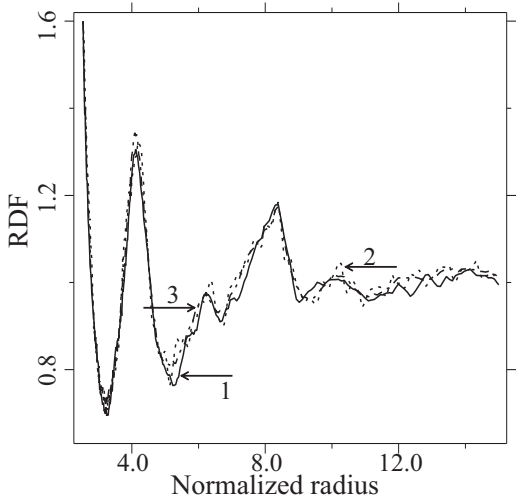
The RDFs $g_{12}(R_{12})$ vs $R_{12}$ at $c=0.6$ estimated by
CRM (1), SCRM (2)
} \ \ \noindent \parbox{7.8cm}{\vspace{0mm}
\hspace{-19mm}
\psfig{figure=figI_13,width=12.0cm}\\
\vspace{-80mm}\tenrm \baselineskip=8pt

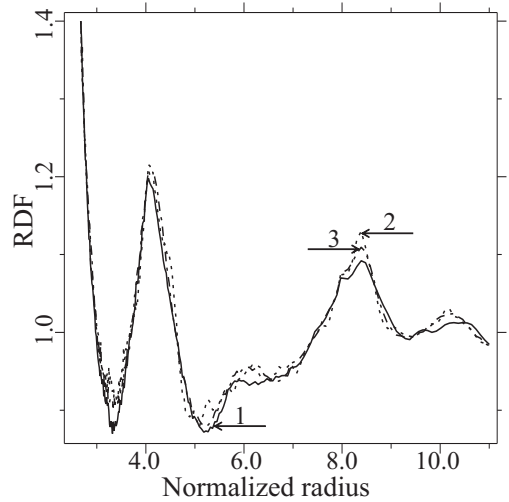
{\rm Fig. 13}. The RDFs $g_{22}(R_{22})$ vs $R_{22}$ at $c=0.6$ estimated by
CRM (1), SCRM (2) }
\vspace{2mm}

\noindent global shaking is not
enough for destruction of locked local arrangements generated by the CRM. So, the difference of $g_{11}(2^+)$ for $M=3$ and $M=60$ equals 12\% while a similar dissimilarity for $M=30$ and $M=60$ does not exceed 0.2\%. Thus, the results of RDFs estimations are statistically representative in both sense the sample size ($N=2000$), statistical stabilization of modeling results ($M=60$), and its smoothness ($P=80$) although the number of big particles $N^{(1)}$ is moderately small ($N^{(1)}=168$ and $N^{(1)}=587$ for $N=2000$ and $N=7000$, respectively).

\medskip
\noindent{\bf 6 Conclusion}
\medskip

{\color{black}Thus the SCRM proposed makes it possible to generate bimodal random structures of spheres for a whole range of particle volume fractions which are primarily homogeneous, highly mixed and protocol independent (in the sense that statistical parameters estimated do not depend on the basic simulated algorithm such as, CRM). Generalization of the proposed algorithm to any number of fractions is straightforward.

It should be mentioned that one of the basic challenges  of the rocket industry is the optimization of size ratios of packs which are usually performed experimentally. Only at the present time this problem can be solved numerically. The problem is reduced to a minimization problem
$f(c^{(1)},\ldots, c^{(N)})\to \min(\max)$ at the constraint $c^{(1)}+\ldots+c^{(N)}\equiv c=c_{\max}$. The objective function $f(c^{(1)},\ldots, c^{(N)})$ can be described as, e.g., either a jamming limit, the values of the RDFs $g_{ij}(a^{(i)}+a^{(j)})$  $(i,j=1,\ldots,N)$ estimated by the parallelized SCRM or the effective properties and stress concentrator factors of local stresses estimated in an accompanying paper by Buryachenko (2012). However, more detailed consideration of this optimization problem is beyond the scope of the current paper and will be analyzed in forthcoming publications by the authors.}

\medskip
\noindent{\bf Acknowledgment:}
\medskip

This work was partially funded by IllinoisRocstar LLC and Micromechanics \& Composites LLC.
VAB also acknowledges the helpful discussions of some problems of spatial statistics with
Professor Dietrich Stoyan.
Both the helpful comments of reviewers and their encouraging recommendations are gratefully acknowledged.

\medskip
\noindent{\bf References}
\medskip

{\baselineskip=9pt
\parskip=1pt

\hangindent=0.4cm\hangafter=1\noindent
{\bf Allen, M.P.; Tildesley, D.J.}  (1987): {\it  Computer Simulations of Liquids}, Oxford Science Publications, Oxford

\hangindent=0.4cm\hangafter=1\noindent{\bf
 Alcaraz, A.N.; Duhau, R.S.; Fern\'andez, J.R.; Harrowell, P.; Miracle, D.B. } (2008):
Dense amorphous packing of binary hard sphere mixtures
with chemical order: The stability of a solute ordered approximation.
 {\it  J. Non-Crystalline Solids},  vol.{ 54}, pp. 3171--3178

 \hangindent=0.4cm\hangafter=1\noindent{\bf
 Amparano, F.E.; Xib, Y.; Roh, Y.--S. } (2000): Experimental study on the effect of aggregate content on fracture behavior of concrete. {\it  Engineering Fracture Mechanics},  vol.{ 67}, 65--84

\hangindent=0.4cm\hangafter=1\noindent{\bf
Aste, T.; Saadatfar, M.; Sakellariou, A.; Senden, T.J. } (2004):
Investigating the geometrical structure of disordered sphere packings.
{\it  Physica},  vol.{ A339}, pp. 16--23

 \hangindent=0.4cm\hangafter=1\noindent{\bf
 Aste, T.; Saadatfar, M.; Senden, T.J, } (2005):  Geometrical structure of disordered sphere packings.
 {\it  Phys. Rev.},  vol.{ E71}, 061302

  \hangindent=0.4cm\hangafter=1\noindent{\bf
 Attard, P. } (1989):   Spherically inhomogeneous fluids. II. Hard-sphere solute in a hard-sphere
solvent. {\it  J. Chem. Phys.},  vol.{ 91}, pp. 3083--3089

\hangindent=0.4cm\hangafter=1\noindent{\bf
Baer, M.R.; Hall, C.A.; Gustavsen, R.L.; Hooks, D.E.; Sheffield, S.A. } (2007):
Isentropic loading experiments of a plastic bonded explosive and
constituents. {\it  J. Appl. Phys.},  vol.{ 101}, 034906

\hangindent=0.4cm\hangafter=1\noindent{\bf
Barrioa, C.; Solana, J.R. } (2005):  Mapping a hard-sphere fluid mixture onto a
single component hard-sphere fluid. {\it  Physica},  vol.{ A 351}, pp. 387--403

\hangindent=0.4cm\hangafter=1\noindent{\bf
Barker, G.C. } (1993):  Computer simulation of granular materials. In.: A. Mehta (Ed.)
{\it  Granular Matter - An Interdisciplinary Approach.}
Springer Verlag, Berlin, 35--84

\hangindent=0.4cm\hangafter=1\noindent{\bf
Banerjee, B.; Adams, D. O. } (2004):
On predicting the effective elastic properties of polymer
bonded explosives using the recursive cell method
{\it  Int. J. Solids Structures},  vol.{ 41}, pp. 481--509

\hangindent=0.4cm\hangafter=1\noindent{\bf
Bennet, C. H. } (1972):  Serially deposited amorphous aggregates of hard spheres.
{\it  J. Appl. Phys.},  vol.{ 43}, pp. 2727--2734.

\hangindent=0.4cm\hangafter=1\noindent{\bf
Berryman, J. G. } (1983):  Random close packing of hard spheres and disks.
{\it  Physical Review},  vol.{ A 27}, pp. 1053--1061.

\hangindent=0.4cm\hangafter=1\noindent{\bf
Binder, K.; Heerman, D. W. } (1997):  {\it  Monte Carlo Simulation in Statistical Physics:
an Introduction.} Springer, Berlin, NY.

\hangindent=0.4cm\hangafter=1\noindent{\bf
van Blaaderen, A.; Wiltzius, P. } (1995):
Real-space structure of colloidal hard-sphere glasses.
{\it  Science},  vol.{ 270}(5239), pp. 1177--1179

\hangindent=0.4cm\hangafter=1\noindent{\bf
Boudreaux, D. S.; Gregor, J. M. } (1977):
Structure simulation of transition-metal-metalloid glasses.
{\it  J. Appl. Phys.},  vol.{ 48}, pp. 152--158.

\hangindent=0.4cm\hangafter=1\noindent{\bf
Buryachenko, V. A. } (2007):  {\it  Micromechanics of Heterogeneous Materials}. Springer, NY.

\hangindent=0.4cm\hangafter=1\noindent{\bf
Buryachenko, V.A.} (2012):
Modeling of random bimodal structures of composites (application to solid propellants):
II. Estimation of effective elastic moduli. {\it Computer Modeling in Engineering \& Sciences.}
(In Press).

\hangindent=0.4cm\hangafter=1\noindent{\bf
Buryachenko, V. A.; Pagano, N. J.; Kim, R. Y.; Spowart, J. E. } (2003):  Quantitative description of random microstructures of composites and their effective elastic moduli. {\it  Int. J. Solids Struct.},
 vol.{ 40}, pp. 47--72.

\hangindent=0.4cm\hangafter=1\noindent{\bf
Cesarano, III J.; McEuen, M. J.; Swiler, T. } (1995):  Computer simulation of particle packing.
{\it  Intern. SAMPE Technical Conf.}  vol.{ 27}, pp. 658--665.

\hangindent=0.4cm\hangafter=1\noindent{\bf
Cheng, Y. F.; Guo, S.J.; Lay, H. Y. } (2000):
Dynamic simulation of random packing of spherical particles.
{\it  Powder Technology},  vol.{ 107}, pp. 123--130.

\hangindent=0.4cm\hangafter=1\noindent{\bf
Chong, J.S.; Christiansen, E.B.; Baer, A.D. } (1971):  Rheology of concentrated suspensions.
 {\it  J. Applied Polymer Science},  vol.{ 15}, pp. 2007--2021

\hangindent=0.4cm\hangafter=1\noindent{\bf
Clarke, A. S.; Willey, J. D. } (1987):  Numerical simulation of the dense random packing
of a binary mixture of hard spheres: Amorphous metals.
{\it  Phys. Rev.},  vol.{ B. 35}, pp. 7350--7356.

\hangindent=0.4cm\hangafter=1\noindent{\bf
Diggle, P.J. } (2003):  {\it  Statistical Analysis of Spatial Point Patterns} (2nd edn).
Academic Press, New York

\hangindent=0.4cm\hangafter=1\noindent{\bf
Donev, A.; Torquato, S.; Stillinger F. H. } (2005):
Pair correlation function characteristics of nearly jammed disordered and ordered hard-sphere packings.
{\it  J. of Computational Physics},  vol.{ 202}, pp. 737--764

\hangindent=0.4cm\hangafter=1\noindent{\bf
Furukawa, K.; Imai, K.; Kurashige, M. } (2000):
Simulated effect of box size and wall on porosity of
random packing of spherical particles.
{\it  Acta Mechanica},  vol.{ 140}, pp. 219--231

\hangindent=0.4cm\hangafter=1\noindent{\bf
Gajdo\v{s}\'ik, J.; Zeman, J.;  Sejnoha, M.} (2006): Qualitative analysis of fiber composite microstructure: Influence of
boundary conditions. {\it Probabilistic Engineering Mechanics}, vol. 21, pp. 317--329

\hangindent=0.4cm\hangafter=1\noindent{\bf
Gallier, S. } (2009):  A stochastic pocket model for aluminum agglomeration in solid
propellants. {\it  Propellants Explos. Pyrotech.},  vol.{ 34}, pp. 97--105

 \hangindent=0.4cm\hangafter=1\noindent{\bf
 Garboczi, E.J.; Bentz, D.P. } (1997):  Analytical formulas for interfacial transition zone properties.
 {\it  Advanced Cement Based Materials},  vol.{ 6}, pp. 99-108

\hangindent=0.4cm\hangafter=1\noindent{\bf
Gavrikov, V.L.; Stoyan, D.} (1995):  The use of marked point processes in
ecological and environmental forest studies.
{\it  Environmental and Ecological Statistics},  vol.{ 2}, pp. 331--344

\hangindent=0.4cm\hangafter=1\noindent{\bf
Gidaspow, D.; Huilin, L. } (1998):
Equation of state and radial distribution functions of FCC particles in a CFB.
{\it  AICHE J.},  vol.{ 44}, pp. 279--293


\hangindent=0.4cm\hangafter=1\noindent{\bf
Hall, P. } (1988):  {\it  Introduction to the Theory of
Coverage Processes.} John Willey \& Sons, NY.

\hangindent=0.4cm\hangafter=1\noindent{\bf
Hansen, J.P.; McDonald, I.R. } (1986):  {\it  Theory of Simple Liquids}. Academic Press, New
York

\hangindent=0.4cm\hangafter=1\noindent{\bf
He, D.; Ekere, N .N. } (2001):
Structure simulation of concentrated suspensions of hard spherical particles.
{\it  AIChE Journal}  vol.{ 47}, pp. 53--59.

\hangindent=0.4cm\hangafter=1\noindent

 \hangindent=0.4cm\hangafter=1\noindent{\bf
 Illian, J.; Penttinen, A.; Stoyan, H.; Stoyan, D. } (2008):
{\it  Statistical Analysis and Modeling of Spatial Point Patterns}.
Willey \& Sons, Chichester.

 \hangindent=0.4cm\hangafter=1\noindent{\bf
 Jerier, J.-F.; Richefeu, V.; Imbault, D.; Donz\'e, F.V. } (2010):
 Packing spherical discrete elements for large scale simulations. {\it
Comput. Methods Applied Mech. Engng},  vol.{ 199}, pp. 1668--1676

\hangindent=0.4cm\hangafter=1\noindent{\bf
Jodrey, W. S.; Tory, M. } (1985):  Computer
simulation of close random packing of equal spheres.
{\it  Phys. Rev.}  vol.{ A32}, 2347.

\hangindent=0.4cm\hangafter=1\noindent{\bf
Kanaun, S.K. } (1990):  Self-consistent averaging schemes in the mechanics of matrix
composite materials. {\it  Mekhanika Kompozitnikh Materialov},  vol.{ 26}, pp. 702--711 (In
Russian. Engl Transl. {\it  Mech Compos Mater},  vol.{ 26}, pp. 984--992)

\hangindent=0.4cm\hangafter=1\noindent{\bf
Kansal, A. R.; Truskett, T. M.; Torquato, S. } (2000):
Nonequilibrium hard-disk packing with controlled
orientational order. {\it  J. Chemical. Phys.},  vol.{ 113}, pp. 4844--4851.

\hangindent=0.4cm\hangafter=1\noindent{\bf
Kanuparthi, S.; Subbarayan, G.; Siegmund, T.; Sammakia, B. } (2009):
The effect of polydispersivity on the
thermal conductivity of particulate
thermal interface materials.
{\it  IEEE Trans. Components Packaging Technologies},  vol.{ 32}, pp. 424--434

\hangindent=0.4cm\hangafter=1\noindent{\bf
Knott, G. M.; Jackson, T. L.; Buckmaster, J. } (2001):
Random packing of heterogeneous propellants.
{\it  AIAA Journal},  vol.{ 39}, pp. 678--686

\hangindent=0.4cm\hangafter=1\noindent{\bf
Kochevets, S.; Buckmaster, J.; Jackson, T.L.; Hegab, A. } (2001):  Random
propellant packs and the flames they support. {\it  AIAA J Propul Power},
 vol.{ 17}, pp. 883--891

\hangindent=0.4cm\hangafter=1\noindent{\bf
K\"onig, D.; Carvajal-{\color{black}Gonzalez}, S.; Downs A. M.; Vassy, J.; Rigaut, J. P. } (1991):
Modeling and analysis of
3-D arrangements of particles by point process with examples of application to biological data obtained by
confocal scanning light microscopy.
{\it  Journal of Microscopy},  vol.{ 161}, pp. 405--433

\hangindent=0.4cm\hangafter=1\noindent{\bf
Kondrachuk, A. V.; Shapovalov, G. G.; Kartuzov, V. V. } (1997):  Simulation modeling of the
randomly nonuniform structure of powders. Two-dimensional formulation of the
problem. {\it  Poroshkovaya Metallurgiya}, (1-2), pp. 111--118 (In Russian. Engl Translation.
{\it  Powder Metall Metal Ceram},  vol.{ 36}, pp. 101--106)

\hangindent=0.4cm\hangafter=1\noindent{\bf
Kotlarchyk, M.; Chen, S.-H. } (1983):
Analysis of small angle neutron scattering spectra from polydisperse interacting colloids.
{\it J. Chem. Phys.}, vol. 79, pp. 2461--2469

\hangindent=0.4cm\hangafter=1\noindent{\bf
Kurita, R.; Weeks, E.R. } (2010):
Experimental study of random-close-packed colloidal particles.
{\it  Physical Rev.},  vol.{ 82}, 011403

\hangindent=0.4cm\hangafter=1\noindent{\bf
Lee, C.H.; Gillman, A.S.; Matou\v{s}, K. } (2011):
Computing overall elastic constants of polydisperse
particulate composites from microtomographic data.
{\it  J. Mech. Phys. Solids},  vol.{ 59}, pp. 1838--1857

\hangindent=0.4cm\hangafter=1\noindent{\bf
Lee, Y.; Fang, C.; Tsou, Y.-R.; Lu, L.-S.; Yang, C.-T. } (2009):
A packing algorithm for three-dimensional convex particles
 {\it  Granular Matter},  vol.{ 11}, pp. 307--315

\hangindent=0.4cm\hangafter=1\noindent{\bf
Lingois, P.; Berglund, L. } (2002):  Modeling elastic properties and volume change
in dental composites. {\it  J. Mater. Sci.},  vol.{ 37}, pp. 4573--4579

\hangindent=0.4cm\hangafter=1\noindent{\bf
Lochmann, K.; Oger, L.; Stoyan, D. } (2006):
Statistical analysis of random sphere packings
with variable radius distribution
{\it  Solid State Sciences},  vol.{ 8}, pp. 1397--1413

\hangindent=0.4cm\hangafter=1\noindent{\bf
Lu, G. Q.; Ti, L. B.; Ishizaki, K. } (1994):  A new algorithm for simulating the random packing of
monosized powder in CIP processes.
{\it  Materials Manufacturing Processes}
 vol.{ 9}, pp. 601--621.

\hangindent=0.4cm\hangafter=1\noindent{\bf
Lubachevsky, B. D.; Stillinger, F. H. } (1990):  Geometric properties of random disk packing.
{\it  J. of Statistical Physics},  vol.{ 60}, pp. 561--583.

\hangindent=0.4cm\hangafter=1\noindent{\bf
Lubachevsky, B. D.; Stillinger, F. H.; Pinson, E. N. } (1991):
Disks vs spheres: Contrasting properties of random packing.
{\it  J. Statistical Phys.},  vol.{ 64}, pp. 501--524.

\hangindent=0.4cm\hangafter=1\noindent{\bf
Luijten, E. } (2006):  Introduction to cluster Monte Carlo algorithms. {\it  Lecture Notes in Physics},  vol.{ 703}, pp. 13-38

\hangindent=0.4cm\hangafter=1\noindent{\bf
Lusti, H.R.; Gusev, A.A.} (2004): Finite element predictions for the thermoelastic
properties of nanotube reinforced polymers. {\it Model. Simul. Mater. Sci. Eng.}, vol. 12,
pp. S107–-S119

\hangindent=0.4cm\hangafter=1\noindent{\bf
Maggi, F.; Bandera, A.; Galfetti, L.; De Luca, L.T.; Jackson, T. L. } (2010):
Efficient solid rocket propulsion for access to space.
{\it  Acta Astronautica},  vol.{ 66}, pp. 1563--1573

\hangindent=0.4cm\hangafter=1\noindent{\bf
Maggi, F.; Stafford, S.; Jackson, T.L.; Buckmaster, J. } (2008):  Nature of packs used in propellant modeling. {\it  Physical Review},  vol.{ E77}, 046107

\hangindent=0.4cm\hangafter=1\noindent{\bf
Markov, K.Z.; Willis, J.R. } (1998):  On the two-point correlation function for dispersions
of nonoverlapping spheres. {\it  Math Models Methods Appl Sci},  vol.{ 8}, pp. 359--377

\hangindent=0.4cm\hangafter=1\noindent{\bf
Massa, L.; Jackson, T.L.; Short, M. } (2003):
Numerical solution of three-dimensional
heterogeneous solid propellants.
{\it  Combust. Theory Modeling},  vol.{ 7}, pp. 579--602

\hangindent=0.4cm\hangafter=1\noindent{\bf
Matou\v{s}, K.; Geubelle, P.H. } (2006):
Multiscale modelling of particle debonding in reinforced
elastomers subjected to finite deformations. {\it  Int.  J. Numerical Methods in Engineering},  vol.{ 65}, pp. 190--223

\hangindent=0.4cm\hangafter=1\noindent{\bf
Matou\v{s}, K.; Inglis H.M.; Gu X.; Rypl, D.; Jackson, T.L.; Geubelle, H.P. } (2007):
Multiscale modeling of solid propellants: From particle packing to failure.
{\it  Composites Science and Technology},  vol.{ 67}, pp. 1694--1708

\hangindent=0.4cm\hangafter=1\noindent{\bf
Matou\v{s}, K.; Lep\v{s}, M.; Zeman, J.; \v{S}ejnoha M.} (2000): Applying genetic
algorithms to selected topics commonly encountered in engineering practice. {\it Computer
Methods in Applied Mechanics and Engineering}, vol. 190, pp. 1629--1650

\hangindent=0.4cm\hangafter=1\noindent{\bf
 McGeary, R.K. } (1961):  Mechanical packing of spherical particles. {\it  J. Am. Ceram. Soc.},  vol.{ 44}, pp. 513--522


\hangindent=0.4cm\hangafter=1\noindent{\bf
Nolan,  G. T.; Kavanagh, P.E. } (1992):  Computer simulation of random packing of hard spheres.
{\it  Powder Technology},  vol.{ 72}, pp. 149--155.

\hangindent=0.4cm\hangafter=1\noindent{\bf
Nolan, G. T.; Kavanagh, P.E. } (1995):  Octahedral configurations of random close packing. {\it  Powder Technology}  vol.{ 83}, pp.  253--258.

\hangindent=0.4cm\hangafter=1\noindent{\bf
Nowak, E.R.; Knight, J.B.; Povinelli, M.L.; Jaeger, H.M.; Nagel, S.R. } (1997):
Reversibility and irreversibility in the packing of vibrated granular material.
{\it  Powder Technology},  vol.{ 94}, pp. 79--83

\hangindent=0.4cm\hangafter=1\noindent{\bf
Oger, L.; Troadec, J. P.; Gervois, A.; Medvedev, N. } (1998):
Computer simulation and tessellations of
granular materials. {\it  {\color{black}Foams} and Emulsions}.
(Eds. N. Rivier, J. F. Sadoc), 527--545. Kluver, Dordrecht.

\hangindent=0.4cm\hangafter=1\noindent{\bf
Okabe, A.; Boots, B.; Sugihara, K. } (1992):  {\it  Spatial Tessellations}. Willey, NY.

\hangindent=0.4cm\hangafter=1\noindent{\bf
Percus, J.K.; Yevick, G.J. } (1958):  Analysis of classical statistical mechanics by means
of collective coordinates. {\it  Phys. Rev.},  vol.{ 110}, pp. 1--13

\hangindent=0.4cm\hangafter=1\noindent{\bf
Picka, J. } (2007):  Statistical inference for disordered sphere packings.  {\it  Arxiv preprint} arXiv: 0711.3035

\hangindent=0.4cm\hangafter=1\noindent{\bf
Pyrz, R. } (2004):  Microstructural description of composites, Statistical methods.
In: B\"ohm H.J. (ed) {\it  Mechanics of microstructured materials}. Springer, Berlin,
Heidelberg, NY, pp. 173--233

\hangindent=0.4cm\hangafter=1\noindent{\bf
Ripley, B.D. } (1977):  Modeling spatial patterns. {\it  J Roy Statist Soc},  vol.{ B39}, pp. 172--212

\hangindent=0.4cm\hangafter=1\noindent{\bf
Ripley, B. D. } (1979):  Tests of ``randomness" for spatial point patterns. {\it  J. R. Statist. Soc.},  vol.{ B 41}, pp. 368--374

\hangindent=0.4cm\hangafter=1\noindent{\bf
Ripley, B.D. } (1981):  {\it  Spatial {\color{black}Statistics}}. John Wiley \& Sons, New York

\hangindent=0.4cm\hangafter=1\noindent{\bf
Rudge, J.F.; Holness, M.B.; Smith, G.C. } (2008):
Quantitative textural analysis of packings of elongate crystals
{\it  Contrib Mineral Petrol},  vol.{ 156}, pp. 413--429

\hangindent=0.4cm\hangafter=1\noindent{\bf
 Saadatfar, M. } (2009):  Computer simulation of granular materials.
{\it  Computing in Science \& Engineering}.,  vol.{ 11}, pp. 66--74

\hangindent=0.4cm\hangafter=1\noindent{\bf
Schaertl, W.; Sillescu, H. } (1994):
Brownian dynamics of polydisperse colloidal
hard spheres: Equilibrium structures and random
close packings.
{\it  J. Statistical Phys.},  vol.{ 77}, pp. 1007-1025

\hangindent=0.4cm\hangafter=1\noindent{\bf
Segurado, J.; Llorca, J. } (2002):  A numerical approximation to the elastic properties of sphere-reinforced composites. {\it  J. Mech. Phys. Solids},  vol.{ 50}, pp. 2107--2121

\hangindent=0.4cm\hangafter=1\noindent{\bf
Shapiro, A.P.; Probstein, R.F. } (1992):  Random packings of spheres and fluidity limits of
monodisperse and bidisperse suspensions. {\it  Phys. Rev. Lett.},  vol.{ 68}, pp. 1422--1425

\hangindent=0.4cm\hangafter=1\noindent{\bf
Sierou, A.; Brady, J.F. } (2002):  Rheology and microstructure in concentrated noncolloidal suspensions.
{\it  J. Rheology},  vol.{ 46}, pp. 1031--1056

\hangindent=0.4cm\hangafter=1\noindent{\bf
Silberschmidt, V. } (2008):  Account for random microstructure in multiscale models.
In: Kwon, Y.W.; Allen, D.H.; Talreja, R.R. (eds)
 {\it  Multiscale Modeling and Simulation of Composite Materials and Structures}. Springer, NY.
pp. 1--35

\hangindent=0.4cm\hangafter=1\noindent{\bf
Silbert, L.E.; Ertas, D.; Grest, G.S.; Halsey, T.C.; Levine, D. } (2002):
Geometry of frictionless and frictional sphere packings.
{\it Physical Review E}, vol. 65, 031304

\hangindent=0.4cm\hangafter=1\noindent{\bf
Stafford, D.S.; Jackson, T.L. } (2010):  Using level sets for creating virtual random packs of non-spherical
convex shapes. {\it  J. Comput. Physics}.  vol.{ 229}, pp. 3295--3315

\hangindent=0.4cm\hangafter=1\noindent{\bf
Steinhauser, O.M.; Hiermaier, S. } (2009):
A review of computational methods in materials science:
examples from shock-wave and polymer physics.
{\it  Int. J. Mol. Sci.},  vol.{ 10}, pp. 5135--5216

\hangindent=0.4cm\hangafter=1\noindent{\bf
Stell, G.; Rirvold, P.A. } (1987):  Polydispersity in fluid, dispersions, and composites:
some theoretical results. {\it  Chem Engng Commun},  vol.{ 51}, pp. 233--260

\hangindent=0.4cm\hangafter=1\noindent{\bf
Stickel, J.J.; Powell, R.L. } (2005):  Fluid mechanics and rheology of dense suspensions.
{\it  Annu. Rev. Fluid Mech.,}  vol.{ 37}, pp. 129--149

\hangindent=0.4cm\hangafter=1\noindent{\bf
Stoyan, D. } (1998):  Random sets: models and statistics. {\it  Int Statistical Rev.},  vol.{ 66}, pp. 1--27

\hangindent=0.4cm\hangafter=1\noindent{\bf
Stoyan, D. } (2000):  Basic ideas of spatial statistics.
{\it  Statistical Physics and Spatial Statistics: the art
of analyzing and modeling Saptial Structures and Patern Formation}.
(Eds. K. R. Mecke, D. Stoyan). Lecture notes in physics, Vol. 554, Berlin, pp. 3--21

\hangindent=0.4cm\hangafter=1\noindent{\bf
 Stoyan D.; Stoyan, H. } (1994):  {\it  Fractals, Random Shapes and Point Fields.
Methods of Geometric Statistics.}
J. Wiley \& Sons., Chichester.

\hangindent=0.4cm\hangafter=1\noindent{\bf
Tanimoto, Y.; Nishiwaki, T.; Nemoto, K.; Ben, G. } (2004):  Effect of filler content on bending properties of dental composites: Numerical simulation with the use of the finite-element method.
 {\it  J. Biomedical Materials Research},  vol.{ 71B}, pp. 188-195

 \hangindent=0.4cm\hangafter=1\noindent{\bf
Throop, G.J.; Bearman, R.J. } (1965):
Numerical solutions of the Percus--Yevick equation for the hard-sphere potential
{\it  J. Chem. Phys.}  vol.{ 42}, pp. 2408--2411

 \hangindent=0.4cm\hangafter=1\noindent{\bf
Tobochnik, J.; Chapin, P.M. } (1988):  Monte Carlo simulation of hard spheres
near random closest packing using spherical boundary conditions. {\it  J Chem Phys},
 vol.{ 88}, pp. 5824--5830

\hangindent=0.4cm\hangafter=1\noindent{\bf
Torquato, S. } (2002):  {\it  Random Heterogeneous Materials:
Microstructure and Macroscopic Properties.}
Springer-Verlag, NY

\hangindent=0.4cm\hangafter=1\noindent{\bf
Torquato, S.; Stell, G. } (1985):  Microstructure of two-phase random media. {\it  J Chem
Phys},  vol.{ 82}, pp. 980--987

 \hangindent=0.4cm\hangafter=1\noindent{\bf
Torquato, S.; Truskett, T.M.; Debenetti, P.G.
} (2000):  Is random close packing of spheres well defined?
{\it  Phys. Rev. Letter},  vol.{ 84}, pp. 2064--2067

 \hangindent=0.4cm\hangafter=1\noindent{\bf
Webb, M.D.; Davis, I.L.} (2006):
Random particle packing with large particle size variations using reduced-dimension algorithms.
{\it Powder technology}, vol. 167, pp. 10–-19

\hangindent=0.4cm\hangafter=1\noindent{\bf
Willis, J. R. } (1978):  Variational principles and bounds for the overall properties
of composites. In: Provan, J.W. (ed), {\it  Continuum Models of Disordered Systems}.
University of Waterloo Press, Waterloo, pp. 185---215

\hangindent=0.4cm\hangafter=1\noindent{\bf
Wissler, M.; Lusti, H.R.; Oberson, C.; Widmann-Schupak, A.H.; Zappini, G.; Gusev, A.A. } (2003):
Non-additive effects in the elastic behavior of dental composites.
{\it  Advanced Engineering Materials},  vol.{ 5}, pp. 113--116

\hangindent=0.4cm\hangafter=1\noindent{\bf
Xu, F.; Aravas, N.; Sofronis, P. } (2008):
Constitutive modeling of solid propellant materials with
evolving microstructural damage.
{\it  J. Mechanics Physics Solids},  vol.{ 56}, pp. 2050--2073

\hangindent=0.4cm\hangafter=1\noindent{\bf
Zinchenko, A.Z. } (1994):  Algorithm for random close packing of spheres with periodic boundary conditions.
{\it  J. Comput. Phys.},  vol.{ 114}, pp. 298--307

\hangindent=0.4cm\hangafter=1\noindent{\bf
Zou, R.P, Xu, J.Q.; Feng, C.L.; Yu, A.B.; Johnston, S, Standish, N. } (2003):  Packing of multi-sized mixtures
of wet coarse spheres. {\it  Powder Technol.},  vol.{ 130}, pp. 77--83

}
\end{document}